\documentclass[useAMS,usenatbib]{mn2e}

\usepackage{graphicx}
\usepackage{amssymb}
\usepackage{bm}

\def\dif{\mathop{}\!\mathrm{d}}

\title[Long-lived asteroids in the 2:1 resonance]%
{The origin of long-lived asteroids in the 2:1 mean-motion resonance with Jupiter}
\author[O. Chrenko et~al.]{
O. Chrenko$^{1}$\thanks{E-mail: chrenko@sirrah.troja.mff.cuni.cz},
M. Bro\v{z}$^{1}$,
D. Nesvorn\'{y}$^{2}$,
K. Tsiganis$^{3}$,
D. K. Skoulidou$^{3}$
\\
$^{1}$Institute of Astronomy, Charles University in Prague, V Hole\v sovi\v ck\'ach 2, CZ--18000 Prague~8, Czech Republic\\
$^{2}$Department of Space Studies, Southwest Research Institute, 1050 Walnut Street, Suite 300, Boulder, C0--80302, USA\\
$^{3}$Department of Physics, Aristotle University of Thessaloniki, GR--54124 Thessaloniki, Greece
}
\begin{document}

\date{Accepted 2015 May 12. Received 2015 May 12; in original form 2015 April 04}

\pagerange{\pageref{firstpage}--\pageref{lastpage}} \pubyear{2015}

\maketitle

\label{firstpage}


\begin{abstract}

The 2:1 mean-motion resonance with Jupiter harbours two distinct groups of asteroids.
The short-lived population is known to be a transient group
sustained in steady state by the Yarkovsky semimajor axis drift.
The long-lived asteroids, however, can exhibit dynamical lifetimes comparable
to $4\,\mathrm{Gyr}$. They reside near two isolated islands of the phase space
denoted A and B, with an uneven population ratio $B/A\simeq10$. The orbits of A-island
asteroids are predominantly highly inclined, compared to island B. The
size-frequency distribution
is steep but the orbital distribution lacks any evidence of a collisional cluster.
These observational constraints
are somewhat puzzling and therefore the origin of the long-lived asteroids
has not been explained so far.

With the aim to provide a viable explanation,
we first update the resonant population and revisit its physical properties.
Using an $N$-body model with seven planets and the Yarkovsky
effect included, we demonstrate that the dynamical depletion of island A is faster,
in comparison with island B. Then we investigate (i) the survivability of primordial resonant asteroids
and (ii) capture of the population during planetary migration, following a recently
described scenario with an escaping fifth giant planet and a jumping-Jupiter instability.
We also model the collisional evolution of the
resonant population over past $4\,\mathrm{Gyr}$.
Our conclusion is that the long-lived group
was created by resonant capture
from a narrow part of hypothetical outer main-belt
family during planetary migration.
Primordial asteroids surviving the migration were probably
not numerous enough to substantially contribute
to the observed population.

\end{abstract}




\section{Introduction}
\label{sec:intro}

The 2:1 mean-motion resonance with Jupiter, hereinafter denoted as J2/1,
is one of the major first-order Jovian resonances intersecting the
main asteroid belt. It is associated with the Kirkwood gap
known as the Hecuba gap \citep{Kirkwood_1867QB741.K59......,Schweizer_1969AJ.....74..779S},
which is commonly considered to be the
borderline separating the outer main belt and the Cybele region.
Its name is derived from (108) Hecuba which is an outer main-belt
asteroid but technically does not exhibit the exact 2:1 commensurability
\citep{Schubart_1964SAOSR.149.....S}.

The 2:1 resonance, which was originally believed to be
depleted of asteroids, harbours several small-sized bodies
as it was realised by observations in the 20th century
(the first resonant asteroid (1362) Griqua was recognized in
1943 \citep{Rabe_1959AJ.....64...53R}).
The follow-up studies based on analytical and semi-analytical methods
\citep*[e.g.][]{Murray_1986Icar...65...70M,Moons_etal_1998Icar..135..458M},
symplectic mapping \citep[e.g.][]{Roig_Ferraz-Mello_1999P&SS...47..653R}
and frequency analysis
\citep[e.g.][]{Nesvorny_Ferraz-Mello_1997A&A...320..672N}
helped to get insight to the internal structure of the resonance
and to investigate the character of various resonant orbits affected
by a complicated interplay between the J2/1 and overlapping
secular \citep{Morbidelli_Moons_1993Icar..102..316M}
and secondary resonances
\citep*{Wisdom_1987Icar...72..241W,Henrard_etal_1995Icar..115..336H}.
In particular, it was discovered
that while the regions of the overlap give rise to a strongly
chaotic behaviour of the asteroidal orbits, there are
two separate regions of quasi-regular motion located
inside the resonant part of the phase space
\citep{Franklin_1994AJ....107.1890F,Mitchenko_Ferraz-Mello_1997P&SS...45.1587M}.
These regions were designated as stable islands A and B
\citep[][see e.g. Fig.~\ref{fig:map.present}]{Nesvorny_Ferraz-Mello_1997A&A...320..672N}.

However, it was also suggested by \cite*{Ferraz-Mello_etal_1998AJ....116.1491F}
that the stability of the islands could have been weakened in the past.
The mechanism responsible for this increase of chaos
was introduced as 1:1 resonance of the J2/1 libration period
with the period of Jupiter-Saturn great inequality (GI). Although
the present value of the GI period is $880\,\mathrm{yr}$, it
was probably smaller when configuration of planetary orbits was
more compact \citep{Morbidelli_Crida_2007Icar..191..158M}
and it slowly increased during Jupiter's and Saturn's
divergent migration. If the value of the GI period was temporarily
comparable with the libration period of the J2/1 asteroids (which is
typically $\simeq420\,\mathrm{yr}$),
the aforementioned resonance would have occurred.

With ongoing improvements of the observation techniques and
sky surveys, there is now a population
of asteroids inside the J2/1 which is large enough
to be analyzed statistically with numerical methods. Long-term integrations
in a simplified model with four giant planets were performed
by \cite*{Roig_etal_2002MNRAS.335..417R} who identified
both short-lived and long-lived asteroids within the observed
population, which consisted of 53 bodies by then. Investigating
resident lifetimes of the resonant asteroids,
\cite{Roig_etal_2002MNRAS.335..417R}
demonstrated that part of the population escapes from the resonance
on time-scales $\sim$$10\,\mathrm{Myr}$, while other asteroids
may have dynamical lifetimes comparable to the age of the Solar
System. The orbits of these long-lived asteroids were embodied
in island B and part of them was also present in its boundaries,
being affected by chaotic diffusion. Surprisingly, the stable island A was
empty.

In a series of papers focused on the 2:1 mean-motion resonance,
\cite{Broz_etal_2005MNRAS.359.1437B} and
\cite{Broz_Vokrouhlicky_2008MNRAS.390..715B}
identified significantly more bodies as resonant.
The latter catalogue contained 92 short-lived and 182 long-lived asteroids.
They also realised that nine of the long-lived asteroids
were located in island A, while the rest was residing in the island B
and its vicinity. They successfully explained the origin of the short-lived
asteroids by a numerical steady-state model, in which the
resonant population was replenished by an inflow of outer main-belt asteroids
driven by the Yarkovsky semimajor axis drift. However, this model fails
in the case of the long-lived population
which was thought to be created by Yarkovsky induced
injection of the nearby Themis family asteroids
that exhibit similar inclinations as the B-island asteroids.
\cite{Broz_etal_2005MNRAS.359.1437B} showed that the objects transported
from Themis are usually perturbed shortly after
entering the resonance and rarely reach the islands.

\cite{Roig_etal_2002MNRAS.335..417R} argued that the steep
size-frequency distribution (SFD)
of the long-lived asteroids, which was inconsistent
with a steady state
\citep{Dohnanyi_1969JGR....74.2531D},
could imply recent collisional origin.
On the contrary, there was no collisional
cluster identified in the orbital distribution 
by \cite{Broz_Vokrouhlicky_2008MNRAS.390..715B},
thus a collisional family would have to be older than $1\,\mathrm{Gyr}$.
Moreover, the presence of the highly inclined A-island
asteroids would require unrealistic ejection velocities,
assuming that the disruption occurred in the more
populated island B.

As there is no self-consistent explanation,
the main goal of this paper is to provide a reasonable
model for the formation and evolution of the long-lived population,
taking into account the paucity of bodies in island A,
the differences in the inclination and the
non-equilibrium size distribution.

The long dynamical lifetimes of the dynamically stable asteroids,
together with the failure of the aforementioned hypotheses,
strongly indicates that their origin may be traced back
to the epoch of planetary migration. Recent advances in migration
theories suggest that
a primordial compact configuration of planetary orbits
and subsequent violent planetary migration might have led to a destabilization
of several regions that are otherwise stable under the current
planetary configuration. The populations
of Trojans and Hildas \citep*{Nesvorny_etal_2013ApJ...768...45N,Roig_Nesvorny_2014DPS....4640001R}
may serve as examples
of such a process: the primordial populations in the 1:1 and 3:2 resonances
with Jupiter were totally dispersed and the observed populations were
formed by resonant capture of bodies, originating either in the outer
main belt or in the transneptunian disc of comets.
It seems inevitable that the 2:1 resonance with Jupiter also undergoes significant
changes of its location, inner secular structure and asteroid
population, during the planetary migration.

The structure of the paper is as follows.
We first use the latest observational data from the databases
AstOrb, AstDyS, WISE and SDSS to update the observed population
in Section \ref{sec:observed_resonant_pop}.
We also briefly review the characterization
of resonant orbits
and describe a method of dynamical mapping.
In Section \ref{sec:effects_of_jj_instability}, we study
whether the planetary migration may cause depletion
or repopulation of the long-lived J2/1 asteroids.
The dynamical models we create are based on simulations with
prescribed evolution of planets
in context of the modern
migration scenario with five giant planets and \textit{jumping-Jupiter
instability} \citep{Nesvorny_Morbidelli_2012AJ....144..117N}.
Section \ref{sec:GI} is focused on dynamical simulations
covering the late stage of planetary migration during which 
the GI period evolution might have influenced the
stability of the long-lived asteroids.
In Section \ref{sec:collisional_models}, we examine
the effects of collisions on the long-lived population.
Finally, Section \ref{sec:conclusions} is devoted to conclusions.


\section{Observed resonant population}
\label{sec:observed_resonant_pop}

In this section, we first describe methods
we use for identification and description of resonant orbits
as well as for dynamical mapping of the resonance.
Our approach is similar to the one used by 
\cite{Roig_etal_2002MNRAS.335..417R}.
We identify resonant orbits in a recent catalogue of
osculating orbital elements
and we study their dynamical lifetimes on the basis of
long-term numerical integrations in a simplified model
with four giant planets only. Our goal is also to survey 
available data for physical properties
of the population, namely the absolute magnitudes, sizes and albedos.
In the last subsection, we
revisit results of Skoulidou et al. (in preparation) in
order to compare the simplified four-giant-planet
model with a more sophisticated framework including
terrestrial planets and the Yarkovsky effect.
Especially, we look for the differences in the dynamical
decay rates in the stable islands A and B.

\subsection{Characterization of resonant orbits}
\label{sec:characterization_res_orbits}

The 2:1 resonance critical angle is defined as:
\begin{eqnarray}
\sigma \equiv 2 \lambda_{\mathrm{J}} - \lambda - \varpi\, ,
\end{eqnarray}
where $\lambda_{\mathrm{J}}$ and $\lambda$ are the mean longitudes of Jupiter
and of an asteroid, respectively, $\varpi$ is the asteroid's longitude of perihelion.
The critical argument of any body trapped inside the J2/1 resonance librates
(quasi-periodically changes on an interval $<2\pi$) with a typical period of about
$420\,\mathrm{yr}$.
On the other hand, the critical argument of the asteroids outside the resonance 
circulates. This provides us a useful tool for the identification of the 
resonant asteroids.

The libration of $\sigma$ is linked to periodic changes
of the osculating semimajor axis $a$, the eccentricity $e$
and the inclination $I$. These changes are coupled together
as described by the adiabatic invariant $N$ of
the asteroid's motion in the circular and
planar restricted Sun--Jupiter--asteroid system
\begin{eqnarray}
N = \sqrt{a}\left(2-\sqrt{1-e^{2}}\cos{I}\right)\, .
\label{eqn:adiab_inv}
\end{eqnarray}
The presence of other planets and variable eccentricity
of Jupiter's orbit give rise to multiple perturbations
in the J2/1 and prevent integrability of the orbits.

None the less, the $a,e,I$ coupling is preserved to a certain degree.
Semimajor axis oscillates around the libration centre,
which is positioned approximately at the exact resonance $a_{\mathrm{res}}\simeq3.27\,\mathrm{AU}$.
The eccentricity and inclination attain their maximal values
when the oscillation of $a$ is at its minimum and vice versa.

An inconvenient consequence of this behaviour is that the standard
averaging methods for the computation of proper elements
\citep{Knezevic_Milani_2003A&A...403.1165K}
do not retain any information about the libration
amplitude (i.e. the proper semimajor axes of all resonant asteroids
approach the value of $a_{\mathrm{res}}$).  
Therefore the proper elements are not the appropriate choice when
studying resonant orbits.

In order to properly characterize the libration amplitude, we use an alternative set of
resonant (or pseudo-proper) elements \citep{Roig_etal_2002MNRAS.335..417R}.
The idea is to record the osculating orbital elements at the
moment when they reach their extremal values during the libration cycle.
These values can be found 
as the intersections with a suitably defined reference plane in the osculating
elements space. A set of conditions determining this plane can be written
as
\begin{eqnarray}
\sigma=0 \wedge \frac{\dif\sigma}{\dif t}>0 \wedge 
\varpi-\varpi_{\mathrm{J}}=0 \wedge \Omega-\Omega_{\mathrm{J}}=0 \, ,
\label{reson}
\end{eqnarray}
where $\Omega$ denotes the longitude of node and subscript $\mathrm{J}$
is used for Jupiter.
The purpose of the conditions for $\varpi$ and $\Omega$ is to
eliminate secular variations of the resonant elements.
The whole set enables the resonant elements to be recorded when the osculating 
semimajor axis $a$ reaches its minimum, the eccentricity $e$ and the inclination
$I$ attain maxima. Note that as a consequence, the J2/1 asteroids are always depicted
on the left-hand side (closer to the Sun) of the libration centre in the resonant
elements space (see e.g. Fig.~\ref{fig:observed.orbdis}).

Moreover, temporal evolution of resonant elements may serve
as the first indicator of the stability. The reason is that
stable orbits exhibit stable libration with only a small variation
around the mean value. Therefore, successive intersections with the reference plane do 
not move significantly and the recorded resonant elements are
nearly exact constants of motion. The situation for the unstable orbits is just the opposite and
the intersections slowly disperse in time. This fact propagates numerically
to the resulting resonant elements, if we compute their standard deviation
as an error of the mean value obtained over a significant period of time
($\sim$$100$ kyr).

Nevertheless, due to higher order perturbations and secular effects, the
above conditions are seldom satisfied exactly and one has to use less
confined criteria in the following form when numerically integrating the orbits:
\begin{eqnarray}
\left|\sigma\right|<5^{\circ} \wedge \frac{\Delta\sigma}{\Delta t}>0 
\wedge \left|\varpi-\varpi_{\mathrm{J}}\right|<5^{\circ}  \, .
\label{cond}
\end{eqnarray}
We use the difference between successive numerical time steps (denoted
as $\Delta\sigma$) rather then the time derivative.
Note that we completely omit the condition for $\Omega$.
This simplification can be compensated by 
verifying that the maximal value of the inclination is reached when
recording the resonant elements.

\subsection{Dynamical maps}
\label{sec:dynmap}
In this section, we summarize our approach to dynamical mapping of the 2:1 resonance
and its secular structure. Our method is verified by comparing
the map computed for the present configuration of planets with the separatrices
and locations of secular resonances derived by \cite{Moons_etal_1998Icar..135..458M}.
An important outcome is our ability to record
global changes inside the resonance when the planetary orbits are reconfigured
and also to examine dynamical stability in various regions
of the resonant phase space.

Our method employs the definition of the resonant elements.
The respective values should change systematically
in the case of unstable resonant orbits, whereas stable
orbits should only oscillate with small variations.
The same behaviour is exhibited
by actions of a dynamical system; time series of their
extremal or mean values are therefore often used for dynamical
mapping
\citep*[e.g][]{Laskar_1994A&A...287L...9L,Morbidelli_1997Icar..127....1M,Tsiganis_2007Icar..186..484T}.

We proceed as follows. First, we divide an investigated part of the phase space
into the boxes of the same size. Let the central coordinates $(a,e,I)$
of each box represent a set of initial resonant elements (this
is accomplished by setting the osculating angular elements so that
condition~(\ref{reson}) is fulfilled). We integrate the initial orbits
over several millions of years and track the evolution of the
resonant elements. Finally, we determine the differences
$\delta a_{\mathrm{r}}$, $\delta e_{\mathrm{r}}$ and $\delta\sin I_{\mathrm{r}}$
between the initial and final (or the last recorded)
resonant elements. These values represent the
dynamical stability of the initial orbit and we 
use them as a characteristics of the entire box.

For the purpose of expressing the total displacement
in the phase space, we define the distance
\begin{eqnarray}
d \equiv \sqrt{\left(\frac{\delta a_{\mathrm{r}}}{\bar{a}_{\mathrm{r}}}\right)^2
+ \left(\delta e_{\mathrm{r}}\right)^2 + \left(\delta \sin{I_{\mathrm{r}}}\right)^2} \, ,
\label{metric}
\end{eqnarray}
where $\bar{a}_{\mathrm{r}}$ denotes the arithmetic mean of the initial and final
resonant semimajor axis. The distance $d$ is similar to the metric
used in the Hierarchical Clustering Method \citep{Zappala_etal_1995Icar..116..291Z}
for family identification.

\begin{figure}[!h]
\centering
\begin{tabular}{c}
\includegraphics[width=84mm]{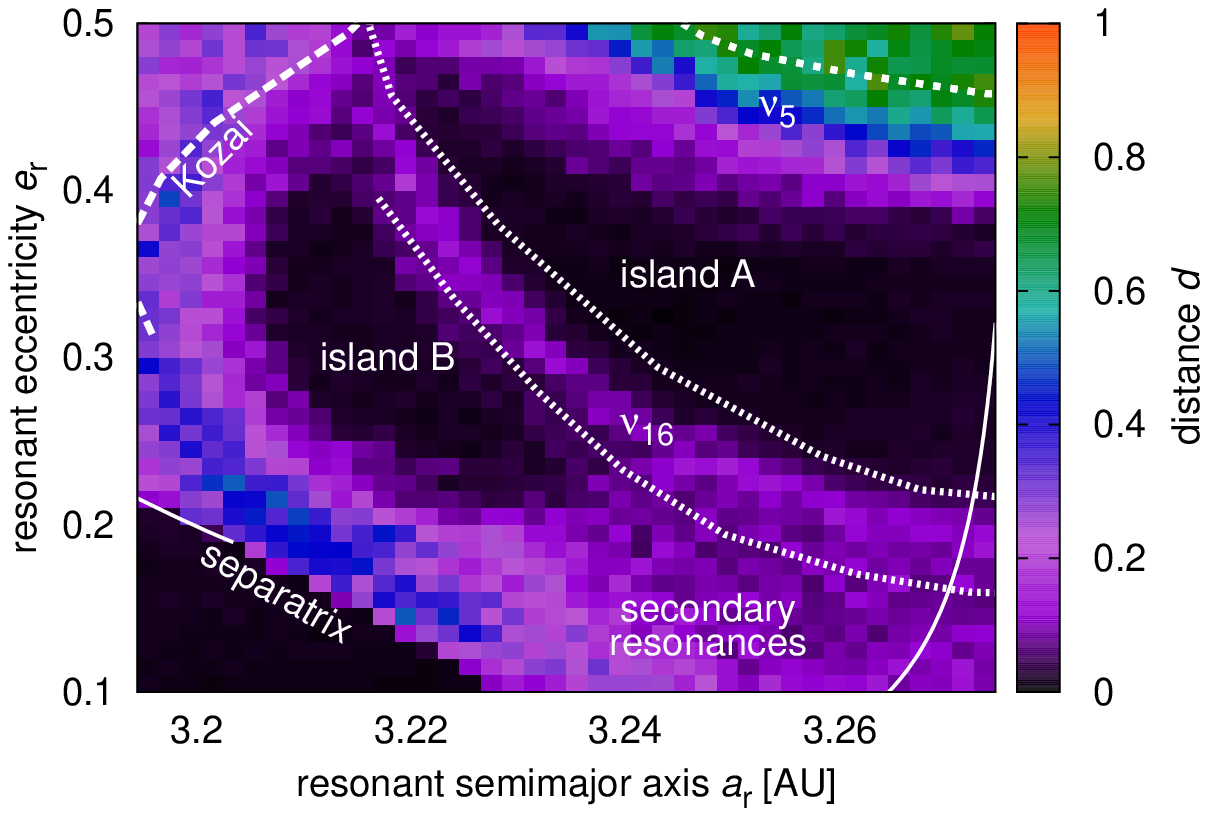} \\
\includegraphics[width=84mm]{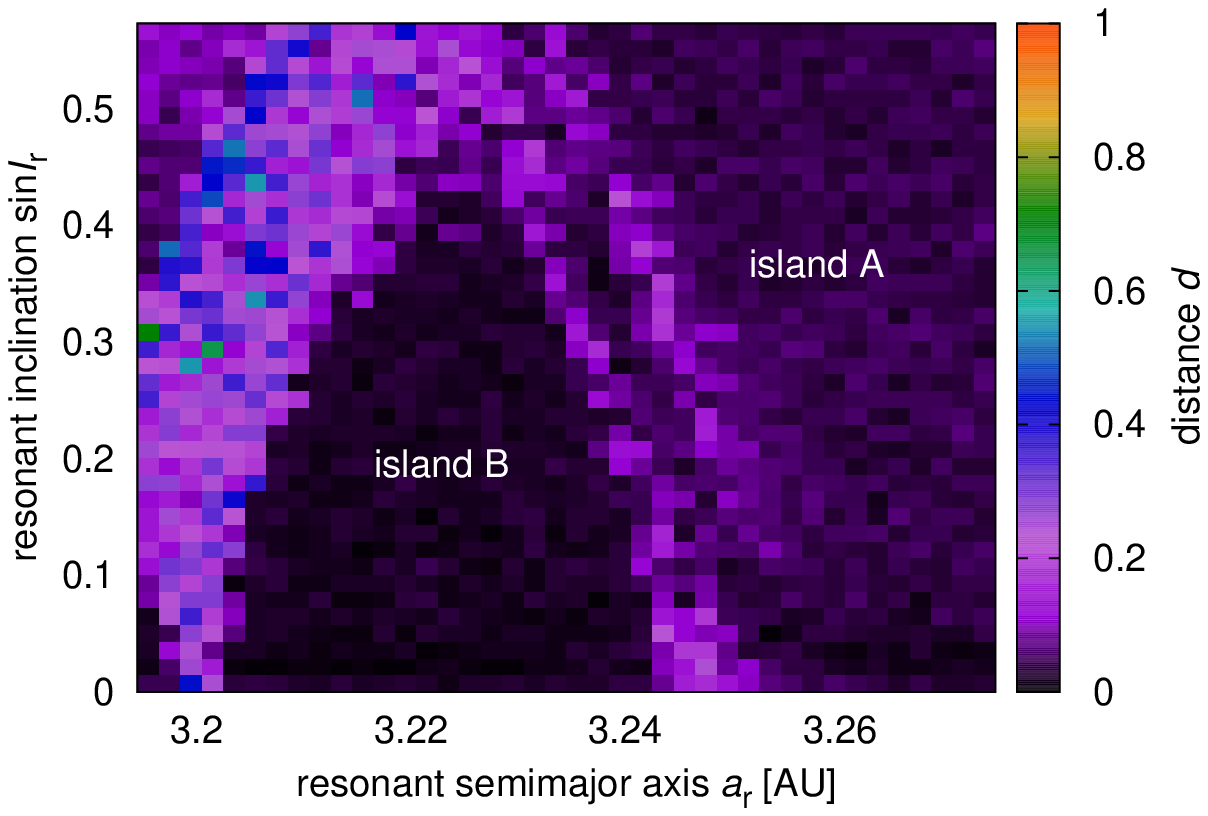} \\
\end{tabular}
\caption{A dynamical map of the 2:1 mean-motion resonance with Jupiter computed
in the present configuration of giant planets. Top panel: the map is plotted
in the resonant semimajor axis $a_{\mathrm{r}}$ vs resonant eccentricity 
$e_{\mathrm{r}}$ plane and it is averaged over five sections in the resonant
inclination $\sin I_{\mathrm{r}}$. The color
coding of boxes represents the average distance in the phase space travelled
by a test particle with initial orbital elements placed within the box. Additionally, 
separatrices and borders of the secular resonances,
which were computed by \protect\cite{Moons_etal_1998Icar..135..458M}, are plotted over the map.
The left solid line corresponds to the separatrix of the J2/1. The 
near-vertical solid line is the libration centre
of the J2/1. The dotted lines indicate the $\nu_{16}$ and $\nu_{5}$ resonance,
and the dashed line represents the Kozai resonance. 
We mark the stable islands denoted A and B and also the region of overlapping
secondary resonances \citep{Roig_Ferraz-Mello_1999P&SS...47..653R}.
Bottom panel: the map plotted
in the resonant semimajor axis $a_{\mathrm{r}}$ vs sine of resonant inclination
$\sin{I_{\mathrm{r}}}$ plane. It was computed for test particles with the resonant
eccentricity $e_{\mathrm{r}}=0.25$.}
\label{fig:map.present}
\end{figure}

In order to verify our method, we constructed a dynamical
map of the 2:1 resonance in the \textit{present} configuration of planets.
We investigated the phase space in the intervals $a\in(3.195,3.275)\,\mathrm{AU}$,
$e\in(0.1,0.5)$, $I\in(0^{\circ},25^{\circ})$ and split them into a grid
of $40\times40\times5$ boxes. With the aim to improve our statistics, we
randomly generated two more test particles in the vicinity of each central
particle (the dispersion of particles is not exceeding $20$ per cent of the box size).
In this way we created the initial conditions for $24,000$ test particles.

In the next step, we integrated the orbits for $t_{\mathrm{span}}=10\,\mathrm{Myr}$.
For all these simulations, we use the symplectic
integrator \textsc{swift} \citep{Levison_Duncan_1994Icar..108...18L} 
with a built-in second-order symplectic scheme
\citep{Laskar_Robutel_2001CeMDA..80...39L}
and with an implementation of digital filters for the computation
of the resonant elements based on criterion~(\ref{cond})
\citep{Broz_etal_2005MNRAS.359.1437B}. This symplectic integrator enables
us to use a time-step $\Delta t = 91.3125\,\mathrm{d}$.
The simulations we perform are simplified:
as we study the outer
main-belt we take into account only the gravitational
interactions with the Sun and the four giant planets. The terrestrial
planets are neglected, except for a barycentric correction applied to 
the initial conditions. We use the Laplace plane to define our initial
frame of reference. We do not consider any non-gravitational
acceleration at this point.

The final step was to
calculate the distance $d$ for each test particle. The average
value $\bar{d}$ for the particles initially placed in the same box was taken
as a measure of the dynamical instability in the box.
Note that there is no temporal measure in this method, i.e.
the map reflects the maximal displacement in each box
during the integration timespan $t_{\mathrm{span}}$
but it does not say how fast did this displacement
occur.

We plot the mean values $\bar{d}$
in the top panel of Figure~\ref{fig:map.present}.
The map is an average projection of all sections
in $\sin I_{\mathrm{r}}$ to the
$\left(a_{\mathrm{r}},e_{\mathrm{r}}\right)$ plane.
It can be compared with several separatrices
derived by \cite{Moons_etal_1998Icar..135..458M}.
It is clear that the map reflects the inner secular
structure of the 2:1 resonance very well:
we can locate both the stable islands A and B,
separated by the $\nu_{16}$ secular resonance.
The borderline at higher values of $e_{\mathrm{r}}$
is formed by the Kozai and $\nu_{5}$ separatrices.
The low-eccentricity region near the libration
centre is affected by presence of multiple
secondary resonances \citep{Roig_Ferraz-Mello_1999P&SS...47..653R}.

The bottom panel of Fig.~\ref{fig:map.present}
displays the projection of our map to the
$\left(a_{\mathrm{r}},\sin I_{\mathrm{r}}\right)$ plane
but this time for a small interval of $e_{\mathrm{r}}$
because the shape of the separatrices and stable islands
strongly depends on $e_{\mathrm{r}}$.
The map demonstrates how the shape of the stable
islands changes with the resonant inclination $I_{\mathrm{r}}$.

Let us conclude that the suitability of the resonant
elements for the dynamical mapping serves as an
independent confirmation that they indeed reflect
the regularity of resonant orbits and they retain
important properties of the proper elements at the same time.

\subsection{Lifetimes of asteroids in the 2:1 resonance}
\label{sec:lifetimes}

Let us now discuss our approach to the identification
and classification of the resonant asteroids.
We numerically propagated the orbits of known numbered and multi-opposition
asteroids in the broad surroundings of the J2/1 to identify those trapped
inside, using a time series of the critical argument $\sigma$.
We extracted the osculating elements of the main-belt objects from the
AstOrb database \citep{Bowell2012} as of 2012 November 15 to set-up the
initial conditions for the first short-term integration.
Our choice of the borders in the osculating $(a,e)$ plane
was $e_{1}=0.45\left(a-3.24\right)/\left(3.1-3.24\right)$ and
$e_{2}=0.5\left(a-3.24\right)/\left(3.46-3.24\right)$.
We ended up with $11,469$ orbits. The corresponding planetary ephemeris were taken
from JPL DE405 \citep{Standish_2004A&A...417.1165S} for the given Julian date. 
We numerically integrated the orbits for $10\,\mathrm{kyr}$ and we recorded the critical
argument $\sigma$ for each asteroid during the simulation. We have found 374
librating asteroids.

The following long-term integration requires a different approach. Our goal is 
to determine the future orbital evolution of the resonant asteroids and
estimate their dynamical lifetime. However, we are limited
by strong chaotic diffusion in the J2/1 resonance.
In other words, even a slight change of the initial conditions can 
significantly alter the orbital evolution. Because the orbital
elements of each observed asteroid are only known with a
finite accuracy, one 
should consider all values within the observational uncertainty as possible
initial conditions to cover all alternatives of future
orbital evolution.

To account for this chaotic behaviour, we first matched the AstOrb data for 
librating asteroids with corresponding AstDyS
uncertainties \citep{Knezevic_Milani_2003A&A...403.1165K}.
The matching failed in four cases\footnote{
These particular bodies have the following designations:
2003~EO31, 2003~HH4, 2006~SE403 and 2006~UF152. The reason for the databases
mismatch might be in different methods used for orbit identification.} only
and hence we discarded these bodies from the J2/1 population.
We then used a pseudo-random generator to create a bundle of 
10 synthetic orbits for each asteroid which are close to its nominal orbit. 
The generated orbital elements fall within the Gaussian distribution
(over $\pm3\sigma$ interval) in the nonsingular osculating element space.
These synthetic test particles are called close clones. In this manner,
we obtained $4,070$ orbits (1 nominal and 10 synthetic 
for each body) which we integrated over $1\,\mathrm{Gyr}$. By this
procedure, we can study several possible realisations of future orbital motion.

Following again \cite{Roig_etal_2002MNRAS.335..417R},
we define the dynamical lifetime $\tau$
as the asteroid's timespan of residence inside the resonance.
Test particles leaving the resonance are usually discarded
due to highly eccentric or inclined orbits, which lead
to planetary crossing or fall into the Sun. We calculated
the median dynamical lifetime $\bar{\tau}$ as the median value of 
the residence lifetimes of close clones.

We divided the J2/1 asteroids into three groups by virtue of their
median dynamical lifetime $\bar{\tau}$ as follows:
\begin{itemize}
  \item{$\bar{\tau}\le70$ Myr: short-lived/unstable/Zulus. Number of identified bodies: 140.}
  \item{$\bar{\tau}\in\left(70,1000\right)$ Myr: long-lived/marginally stable/Griquas. Number of identified bodies: 106.}
  \item{$\bar{\tau}\ge1$ Gyr: long-lived/stable/Zhongguos. Number of identified bodies: 124.}
\end{itemize}
For reference, the numbers of resonant asteroids identified in
the previous paper
\cite{Broz_Vokrouhlicky_2008MNRAS.390..715B}
were 92 short-lived and
182 long-lived asteroids.

We used the same long-term integration of the resonant orbits to
construct orbital distribution of asteroids in the J2/1. 
We calculated average values of the resonant elements for each
asteroid and each close clone over the time interval of $1$ Myr;
then we computed the arithmetic
mean for the asteroid and its close clones together.
We define the standard deviation of the resonant elements
as the uncertainty of the arithmetic mean\footnote{
Let us note that the standard deviations of 
the mean resonant elements will be slightly overestimated.
The reason for this arises from our approach to the generation of
initial conditions because the initial elements of the close clones 
are generated \textit{randomly} thus resulting in uncorrelated sets
of quantities even though they should be correlated, as described
by the correlation matrices available in orbital elements databases.}.

\begin{figure}
\centering
\begin{tabular}{c}
\includegraphics[width=84mm]{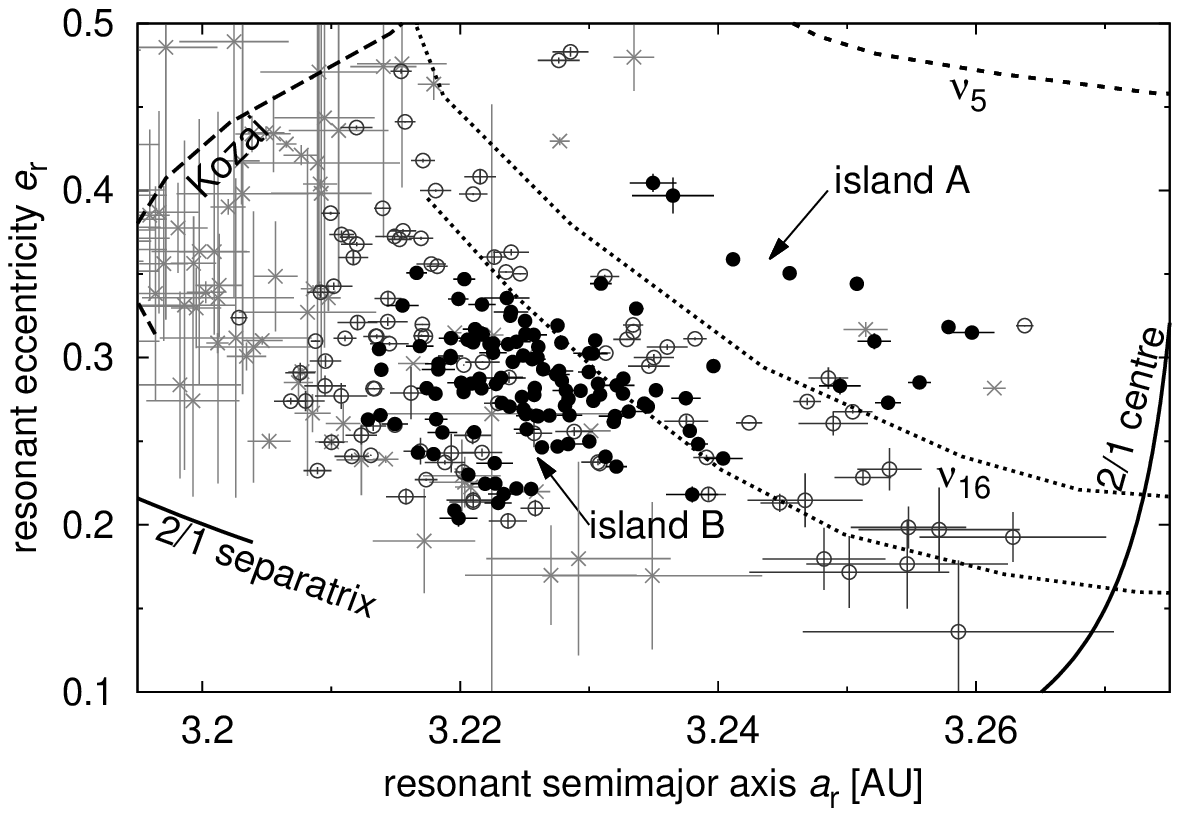} \\
\includegraphics[width=84mm]{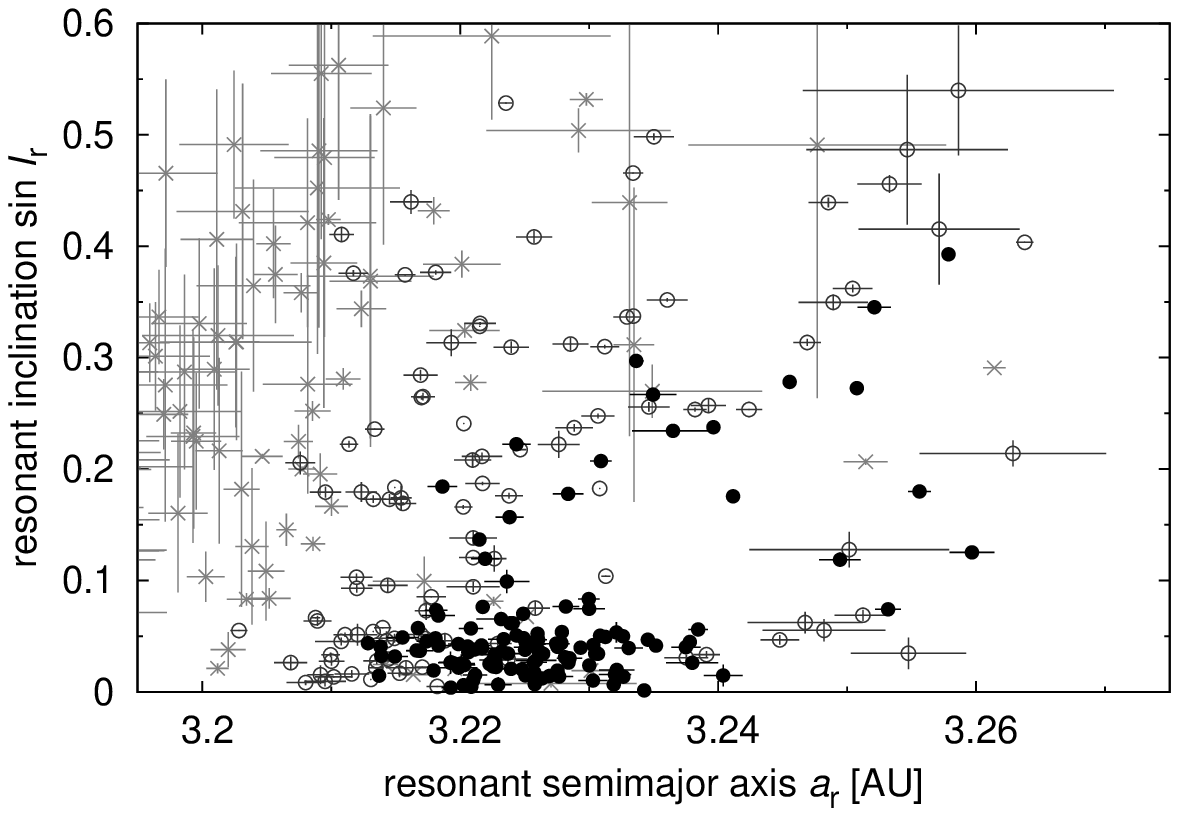} \\
\end{tabular}
\caption{Orbital distribution of the J2/1 asteroids in the resonant
semimajor axis $a_{\mathrm{r}}$ vs the resonant eccentricity $e_{\mathrm{r}}$ plane (top) and in
$a_{\mathrm{r}}$ vs sine of the resonant inclination $\sin{I_{\mathrm{r}}}$
plane (bottom).
The symbols correspond to the dynamical lifetime of each body: filled circles denote
dynamically stable Zhongguos, empty circles marginally stable Griquas,
crosses unstable Zulus. The error bars indicate standard deviations
of the computed orbits. We show the same borders and structures as in Fig.~\ref{fig:map.present}.
Only a relatively small part of the unstable population is depicted.}
\label{fig:observed.orbdis}
\end{figure}

We depict the resulting mean resonant elements as projections
to the $\left(a_{\mathrm{r}},e_{\mathrm{r}}\right)$
and $\left(a_{\mathrm{r}},\sin{I_{\mathrm{r}}}\right)$ planes
in Fig.~\ref{fig:observed.orbdis}.
The orbital distribution exhibits similar properties
as in the previous studies. Zhongguos reside at
the very centre of the stable islands, $11$
Zhongguos in island A and $113$ in island B. The majority of
B-island Zhongguos have low resonant inclinations
while A-island Zhongguos are more inclined.
Griquas partially overlap with the Zhongguo group, but
they either reside on orbits with lower semimajor axis
or higher inclination than Zhongguos. There is no clear
separation between the orbits of Griquas and Zhongguos.
It is therefore questionable whether the division
of the long-lived population into these two groups
has a solid physical foundation, apart from the
dynamical lifetime. The orbital position
of Griquas with respect to Zhongguos indicates
that Griquas might be chaotically diffusing
part of Zhongguos which have haphazardly entered
the periphery regions of the stable islands.

Since we focus on the long-lived population, a great part of
the short-lived bodies is not displayed in Fig.~\ref{fig:observed.orbdis}
(they are located outside the range of the plot).
The depicted short-lived asteroids are chaotically
drifting towards the Kozai separatrix.

\subsection{Albedo, colour and size-frequency distributions}
\label{sec:distributions}

In order to study the basic physical properties
of the J2/1 population, we first searched the WISE database.
We were able to
extract the visual geometric albedos $p_{\mathrm{V}}$ and 
the effective diameters $D$, inferred from NEATM thermal models by
\cite{Masiero_etal_2011ApJ...741...68M},
for 44 asteroids inside the J2/1 (out of 370).

\begin{figure}
\centering
\includegraphics[width=75mm]{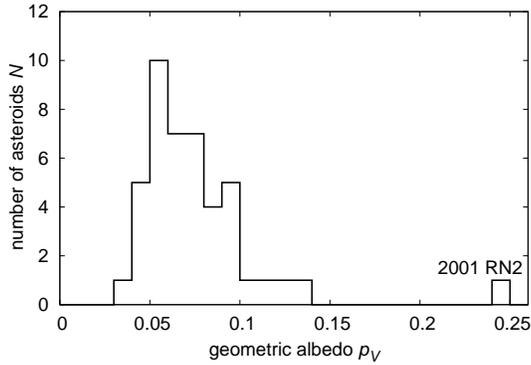}
\caption{A histogram of the albedo distribution among the J2/1 asteroids:
the visual geometric albedo $p_{\mathrm{V}}$ vs the number of asteroids $N$.
The data for 44 asteroids were available in
\protect\cite{Masiero_etal_2011ApJ...741...68M}.}
\label{albedo.histo}
\end{figure}

The data obtained allow us to examine the albedo
distribution in the resonance (Fig.~\ref{albedo.histo}).
The lowest albedo is $\left(0.034\pm0.002\right)$
while the largest value is $\left(0.24\pm0.03\right)$. 
This relatively large value corresponds to asteroid 2001~RN2
and indicates that this object likely belongs to the S taxonomic type.
The majority of the asteroids ($98$ per cent) have albedo lower or
equal to $\left(0.136\pm0.004\right)$.
The shape of the albedo distribution is typical for the outer main-belt
region where C-types dominate
\citep[e.g.][]{Demeo_Carry-2013Icar..226..723D}.

As the next step, we investigated the diameters of asteroids. For the 44 cases
we simply use the value $D$ and its uncertainty.
For the rest of the asteroids we calculate their approximate diameter 
using the following relation \citep{Harris_1998Icar..131..291H}:
\begin{eqnarray}
D = \frac{1329}{\sqrt{p_{\mathrm{V}}}}10^{-\frac{H}{5}} \, ,
\end{eqnarray}
where we insert the absolute magnitude $H$ from the AstOrb catalogue
and we use the mean albedo $\bar{p}_{\mathrm{V}} = \left(0.08 \pm 0.03\right)$.
The standard deviation of~$D$
is then computed as the propagated uncertainty of both $H$ and~$\bar{p}_{\mathrm{V}}$.

\begin{figure}
\centering
\includegraphics[width=84mm]{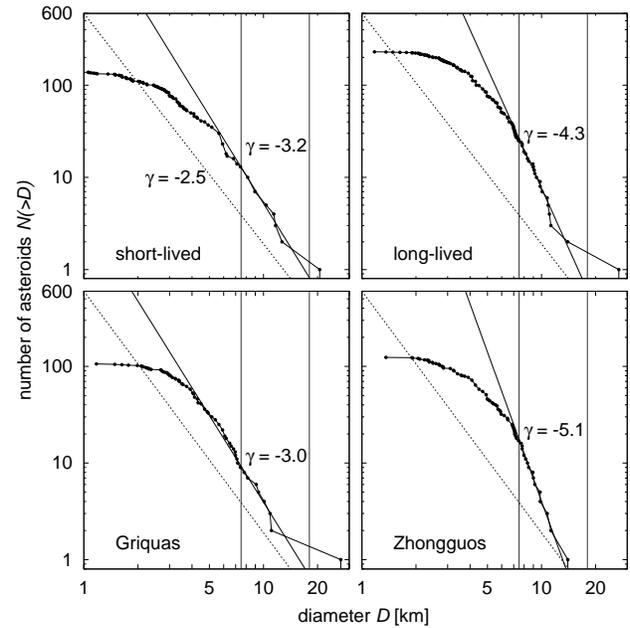}
\caption{Size-frequency distributions (SFDs) of individual dynamical
groups residing inside the J2/1: the diameter $D$ vs the cumulative
number $N\left(>D\right)$ of asteroids larger than $D$. A logarithmic
scaling is used in the plot. The distributions correspond to the short-lived
population (top left) and long-lived population (top right), which
is then divided to marginally stable Griquas (bottom left) and
stable Zhongguos (bottom right).
The steep part of each distribution is approximated with a
power-law $N\left(>D\right) \propto D^{\gamma}$.
The nominal borders of the fitted regions (ranging from $7.5\,\mathrm{km}$ to $18\,\mathrm{km}$)
are shown by vertical lines. The value of the slope $\gamma$
is also given. We also plot stationary Dohnanyi-like
slope ($\gamma=-2.5$) for comparison.
}
\label{fig:sfds}
\end{figure}

We construct the cumulative size-frequency distributions (SFDs) of the resonant population
and individual groups, as shown in Fig.~\ref{fig:sfds}. We fit the steep part of the
distribution with a power-law function $N\left(>D\right) \propto D^{\gamma}$ to
estimate the slope parameter $\gamma$ for further comparison. The value of~$\gamma$
clearly depends on the chosen interval of diameters over which we approximate the
SFDs with the power law. We set the nominal fitted interval as $D\in\left(7.5,18\right)\,\mathrm{km}$.
The SFDs start to bend at the lower limit of this nominal interval
except for SFD of Griquas which remains steep up to $D\simeq4\,\mathrm{km}$.
We also record the variation of the slope $\gamma$
by slightly changing the fitted range of diameters.
The results are summarized in Table~\ref{tab:slopes}.
Unlike the results of \cite{Broz_etal_2005MNRAS.359.1437B},
the SFD of updated Griquas is not shallower
than a Dohnanyi-like SFD with $\gamma=-2.5$
\citep{Dohnanyi_1969JGR....74.2531D}.
In Section~\ref{sec:collisional_models},
we shall use the observed SFDs for comparison
with the outcomes of our collisional models.

\begin{table}
	\centering
	\begin{tabular}{ccc}
		\hline
		group & nominal $\gamma$ & variation of $\gamma$ \\
		\hline
		short-lived & $-3.2$ & $\left(-2.5,-3.7\right)$ \\
		long-lived & $-4.3$ & $\left(-3.7,-5.1\right)$ \\
		Griquas & $-3.0$ & $\left(-3.0,-3.3\right)$ \\
		Zhongguos & $-5.1$ & $\left(-3.9,-5.1\right)$ \\
		\hline
	\end{tabular}
	\caption{The slopes $\gamma$ resulting from the method of least squares
		which we applied to fit the SFDs of dynamical groups with a power law
		function $N\left(>D\right) \propto D^{\gamma}$. The middle column shows
		the result of the fit in the nominal interval of diameters
		$D\in\left(7.5,18\right)\,\mathrm{km}$, the third column reflects
		the variation of $\gamma$ if the limits of the interval are shifted.}
	\label{tab:slopes}
\end{table}

\begin{figure}
	\centering
	\includegraphics[width=84mm]{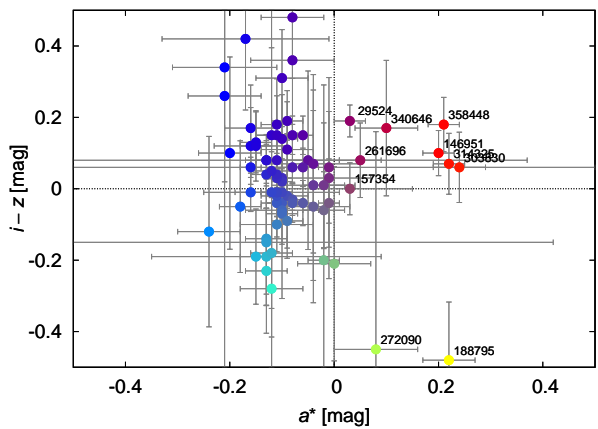}
	\caption{The optimized colour $a^{\star}$ vs the $i-z$ colour diagram
		of the J2/1 resonant asteroids found in the SDSS MOC catalogue.
		Error bars represent the standard deviations of the displayed
		quantities.
		The outliers with $a^{\star}>0$ are labeled with corresponding 
		catalogue numbers.
	}
	\label{fig:colours}
\end{figure}

Finally, we also searched the SDSS MOC catalogue \citep{Parker_etal_2008Icar..198..138P}.
We found astrometric and photometric data
for 81 resonant bodies
for which we constructed a colour-colour diagram
(see Fig.~\ref{fig:colours}). The principal component denoted as $a^{\star}$
is defined on the basis of measurements
in filters $r$, $i$ and $g$ as $a^{\star}=0.89\left(g-r\right)+0.45(r-i)-0.57$
and it can be used to distinguish C-complex ($a^{\star}<0$) and S-complex ($a^{\star}>0$)
asteroids. The distribution in the diagram is typical for
outer main-belt C-type asteroids, but one can also see
several outliers with $a^{\star}>0$. We did not found
any significant relation between the colours and the
orbital distribution.

\subsection{The long-term stability of the islands A and B}
\label{sec:islands_decay}
The observed resonant population has likely evolved over
$\mathrm{Gyr}$-long time-scales. Hence, it is important
to assess its dynamical stability over similarly long time-spans.
This problem was recently revisited by \cite*{Skoulidou_etal_2014}.
Here we will use the main results of this study that are relevant
for our work; a complete dynamical study will be presented in a
different paper (Skoulidou et al. in preparation). 

When studying the outer asteroid belt, we typically ignore the
weak perturbative effects of the terrestrial planets, as this
speeds up the numerical propagation. \cite{Skoulidou_etal_2014} showed
that this is not a safe option, when studying the 2:1 resonant population
for time-spans longer than $1\,\mathrm{Gyr}$, as numerous weak
encounters with Mars
actually induce small-scale chaotic variations. These variations build-up
slowly with time and eventually assist in breaking the phase-protection
mechanism that would otherwise prevent asteroids on moderately eccentric
orbits ($e\sim 0.4$) from encountering Jupiter on this time-span. Moreover,
when the Yarkovsky effect is added, asteroids with $D<20\,\mathrm{km}$
can escape from the resonance more efficiently. This is because,
as also shown in \cite{Broz_Vokrouhlicky_2008MNRAS.390..715B},
the combined action of resonance and slow
Yarkovsky drift forces the asteroid orbits to slowly develop higher
eccentricities as a consequence of adiabatic invariance. The computations
presented in \cite{Skoulidou_etal_2014} show that, in a physical model
that takes into account the aforementioned phenomena, the 2:1 population
would decay roughly exponentially in time, with an e-folding time of
order $1\,\mathrm{Gyr}$.

\begin{figure}
        \centering
        \includegraphics[width=84mm]{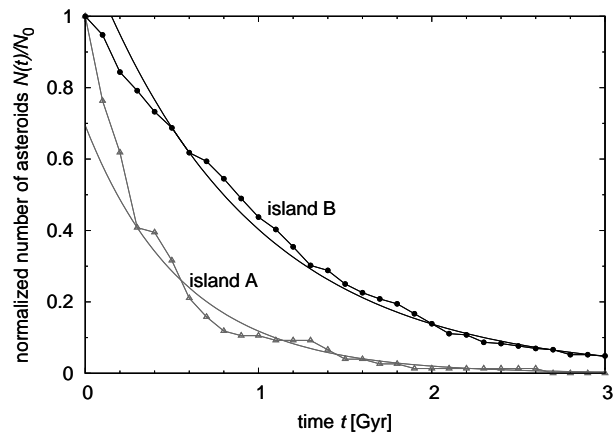}
	\caption{Time evolution of the number of resonant objects,
		$\left(N\left(t\right)/N_{0}\right)$, normalized to the
		initial value. This is the result of a $3\,\mathrm{Gyr}$-long
	simulation, within the framework of the 7-planets model and including
        Yarkovsky-induced drift in $a$. The two curves refer to the two islands
	(A=open triangles and B=solid circles). Superimposed
	are the least-square
	fitted exponentials, yielding e-folding times of
	$\tau_{\mathrm{A}}=\left(0.57\pm 0.02\right)\,\mathrm{Gyr}$
	and $\tau_{\mathrm{B}}=\left(0.94\pm 0.02\right)\,\mathrm{Gyr}$, respectively.}
        \label{fig:longterm_stability}
\end{figure}

In Fig.~\ref{fig:longterm_stability} we present the results of a
similar simulation that
spanned $3\,\mathrm{Gyr}$, in a model that contained the seven major planets
(Venus to Neptune; `7pl' model). This simulation treated the 2:1
population, as was defined by \cite{Broz_Vokrouhlicky_2008MNRAS.390..715B}.
Also, an
approximate treatment of the Yarkovsky acceleration was included
in the equations of motion, assuming a constant drift rate in
semimajor axis equal to
$2.7\times 10^{-4}D^{-1}\,\mathrm{AU}\,\mathrm{Myr}^{-1}$. For each
object its diameter $D$ was estimated, using its catalogued $H$ value
and assuming an albedo of $0.06$--$0.08$, and a value of
either $0^{\circ}$ or $180^{\circ}$ obliquity was randomly assigned.
The population was found to decay
exponentially in time, according to what was described in the
previous paragraph.

However, what is more interesting here is that the two
sub-populations that are contained in the two quasi-stable resonant
islands are not diffusing out at the same rate.
Fig.~\ref{fig:longterm_stability} shows two
sets of curves (one for each island): clearly island A is depleted
faster than island B. Fitting exponentials on the data, we get
the corresponding e-folding times:
$\tau_{\mathrm{A}}=\left(0.57\pm0.02\right)\,\mathrm{Gyr}$ for island A
and $\tau_{\mathrm{B}}=\left(0.94\pm0.02\right)\,\mathrm{Gyr}$ for island B.
This uneven depletion
is an important dynamical property of the resonance and is likely one
of the reasons behind the observed A/B asymmetry.
The importance of this observation will become
more evident in the following sections.  

\newpage
\section{Effects of the jumping-Jupiter instability}
\label{sec:effects_of_jj_instability}
We investigate the role of the jumping-Jupiter instability
on the depletion and eventual repopulation
of the J2/1 in this section. We study survival of a hypothetical
primordial population, and resonant capture from the outer main belt, both in 
the fifth giant planet scenario \citep{Nesvorny_Morbidelli_2012AJ....144..117N}.
Our simulations cover only the instability phase itself, demonstrating at first
which process is plausible.

\subsection{Simulations with prescribed migration}
\label{sec:sims_with_prescribed_mig}
We adopt the technique suggested by 
\cite{Nesvorny_etal_2013ApJ...768...45N}
for simulating the jumping-Jupiter instability. We modified
the \textsc{swift\_rmvs3} integrator \citep{Levison_Duncan_1994Icar..108...18L},
so that the evolution
of massive bodies is given by a prescribed input file
and the evolution of test particles is computed by the 
standard symplectic algorithm. This method
allows us to exclude the transneptunian disc of planetesimals from
our integrations, because its damping and scattering effects are
no longer needed; planets evolve `the way they are told' by the
input file. Moreover, the exact progress of the migration scenario
is then exactly reproducible.

\begin{figure}
\centering
\includegraphics[width=84mm]{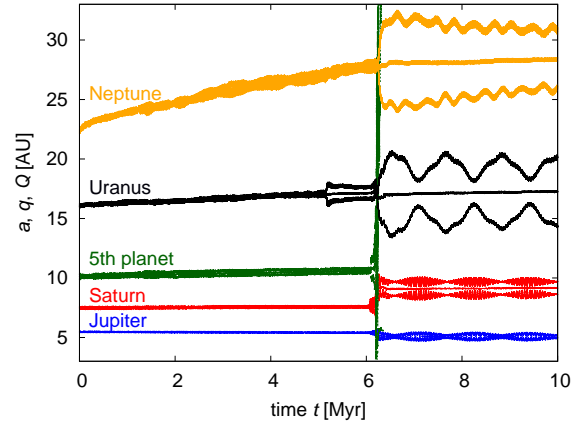}
\caption{Orbital evolution of giant planets in the fifth giant planet
	scenario, adopted from \protect\cite{Nesvorny_Morbidelli_2012AJ....144..117N},
	during the jumping-Jupiter instability,
as it was reproduced by our modified integrator. We plot the time
$t$ vs the semimajor axis $a$, the pericentre $q$ and the apocentre $Q$.
Each evolutionary
track is labeled with the name of the corresponding giant planet.}
\label{fig:mig_scenario}
\end{figure}

The evolution of migrating planets is prescribed by
the fifth giant planet scenario, developed in \cite{Nesvorny_Morbidelli_2012AJ....144..117N}
(see Fig.~\ref{fig:mig_scenario}). This scenario has already been proved reliable
in terms of reproducing the orbital architecture
of planetary orbits, the period ratios and secular
frequencies,
and several distributions of minor solar-system bodies.
The prescribed input file is the result of a complete self-consistent simulation
with five giant planets and a massive transneptunian disc of planetesimals
($M_{\mathrm{disc}}=20\,M_{\mathrm{Earth}}$).
We use only a 10 Myr portion of the simulation, containing the jumping-Jupiter phase.
The input data sampling is $\Delta t_{\mathrm{input}}=1\,\mathrm{yr}$
which is sufficient for our purpose.

Our integrator transforms the input data to the Cartesian coordinates
(i.e. orbital elements to positions and velocities)
and interpolates them according to the integration time step $\Delta t = 0.25\,\mathrm{yr}$.
The interpolation is done by a forward/backward drift along Keplerian ellipses,
starting at input data point preceding/forthcoming to required
time value. The results of both drifts are averaged using the weighted mean,
where the weight depends on temporal distance between the data points and the
required time value.

\subsection{Capture from the outer main belt}
\label{sec:capture}
We investigated a possibility that the resonant asteroids
were captured from outer main-belt orbits near the J2/1
during Jupiter's jump. 
We present a simulation of the capture in this section.

\paragraph*{Initial conditions.}

Initially, we distributed $5,000$ test particles uniformly over the region
which is about to be covered by the moving 2:1 resonance.
This can be estimated easily by Kepler's third law since
we know the evolution of Jupiter's semimajor axis a priori.
Our choice of the semimajor axes distribution was
$a\in(3.06,3.32)\,\mathrm{AU}$.
The distribution of eccentricities and inclinations was also chosen as uniform
and corresponds to the extent of a moderately excited main belt: $e\in(0,0.35)$, $I\in(0^{\circ},15^{\circ})$.
The angular elements were randomly distributed
over $(0^{\circ},360^{\circ})$ interval.

\paragraph*{Evolution of test particles.}

We recorded the time series of resonant elements of the test particles
during the integration
and processed them off-line using the Savitzky--Golay smoothing filter
\citep{Press_etal_2007}
with $0.1\,\mathrm{Myr}$ range of the running window
and a second order smoothing polynomial.

\begin{figure*}
        \centering
        \includegraphics[width=0.82\textwidth]{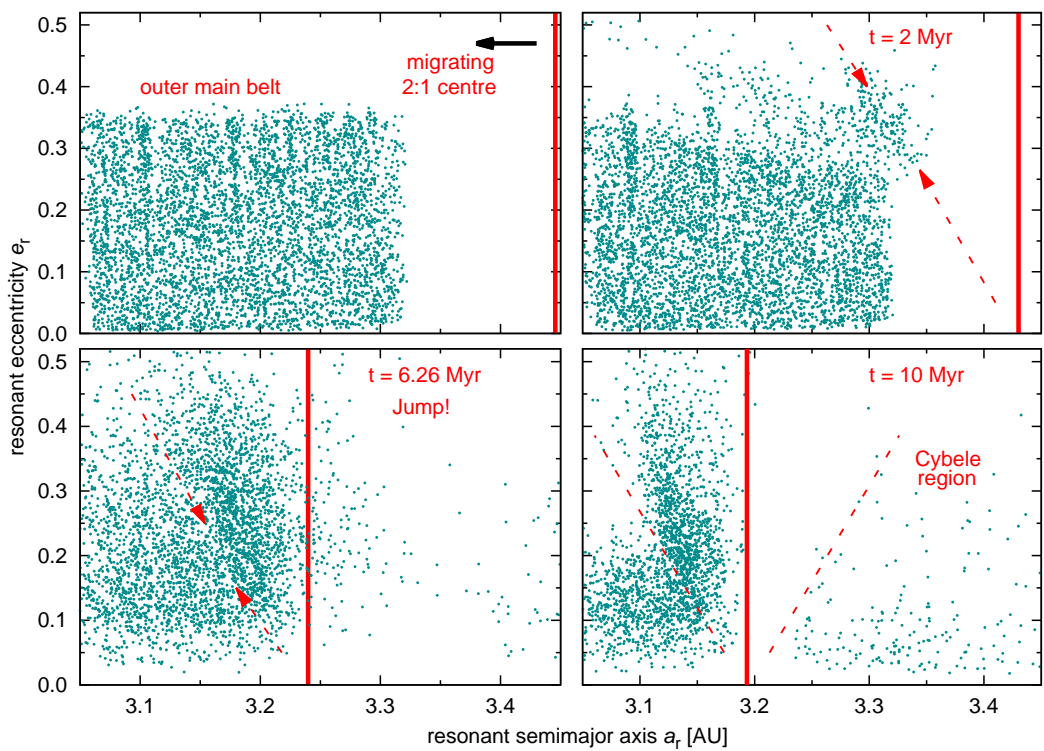} \\
        \caption{A simulation of the resonant capture from the outer main belt
		in the fifth giant planet scenario \protect\citep{Nesvorny_Morbidelli_2012AJ....144..117N}.
                We plot the evolution of test particles in the resonant
                semimajor axis $a_{\mathrm{r}}$ vs resonant eccentricity $e_{\mathrm{r}}$
                plane. The vertical line indicates an
                approximate location of the 2:1 resonance centre which migrates inwards together with
                Jupiter. The integration time is shown
		in three of panels ($t=0$, $2$, $6.26$ and $10\,\mathrm{Myr}$)
		and corresponds to the timeline in Fig.~\ref{fig:mig_scenario}.
		The dashed arrows and lines indicate an approximate extent
		of the libration zone in case of the migrating and stable resonance, respectively.
                From top to bottom and left to right, the plots show: the
                initial conditions, the relaxed population of test particles prior to jump, the state during
                the jumping-Jupiter instability and the final state.
                Note that all test particles are depicted in terms of resonant elements
                for simplicity (even the non-resonant orbits).
 	        }
        \label{fig:capture_evol}
\end{figure*}

The result of our simulation is shown in Fig.~\ref{fig:capture_evol}.
Note that for the sake of simplicity, we use the resonant elements for \textit{all}
particles, even for those not trapped inside the J2/1\footnote{ 
The resonant elements of the non-resonant asteroids do not have
any special physical meaning, but they correspond to the extremal
secular variations of osculating elements as given by equation~(\ref{cond}).
}.
Shortly after the onset of migration,
the test particles still retain the uniform character
of the initial distribution. A few minor mean-motion resonances can
be identified. The apparent gap between the resonance centre
and the synthetic population is intentional, because we do not want
any of the test particles to be initially placed inside the resonance.

After 2 Myr, the relaxation
processes start to take place. The 2:1 resonance
changes its position with respect
to the initial one and starts to perturb several eccentric orbits at the outskirts of
the synthetic population.
All of the perturbed asteroids fall into the short-lived unstable zone
\citep[discussed in][]{Broz_etal_2005MNRAS.359.1437B}.
The minor mean-motion resonances begin to pump the eccentricity of
the test particles residing inside them.

At $6.26\,\mathrm{Myr}$, an instantaneous discontinuity, 
known as \textit{Jupiter's jump}, occurs. The 2:1 resonance, co-moving with jumping
Jupiter, changes its location suddenly and a large number of test particles
flow through the libration zone, while others are excited and ejected,
or left behind in the emerging inner Cybele region.

Finally at $10\,\mathrm{Myr}$, a population
between the left-hand separatrix and the libration centre is stabilised, as the
result of the resonant capture. The bodies outside the left-hand
separatrix are simply remnants of the initial population of test particles.
A few bodies between the libration centre and the right-hand separatrix
are in fact outside the 2:1 resonance, as well as the asteroids
in the Cybele region.

\paragraph*{Captured population.}

Instead of another long-term integration,
we use the rather efficient dynamical mapping introduced
in Section~\ref{sec:dynmap} to check the stability
of the captured population in the
post-migration configuration of planets.
This allows us to identify the counterparts
of the stable islands A and B in the phase
space for this planetary configuration.
Hence, we can trace the long-lived candidates of our captured bodies simply by selecting
particles with resonant elements falling within the range of the islands.

\begin{figure}[!h]
	\centering
	\begin{tabular}{c}
		\includegraphics[width=84mm]{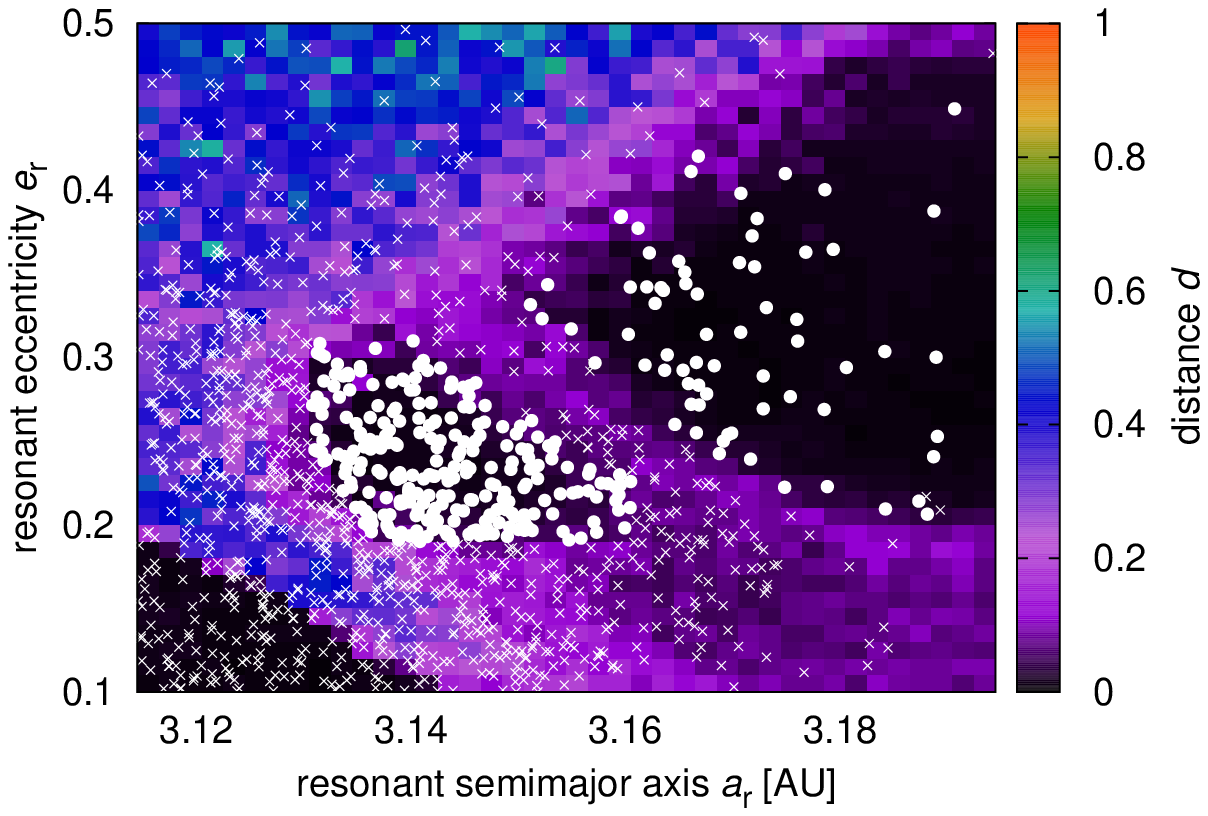} \\
		\includegraphics[width=84mm]{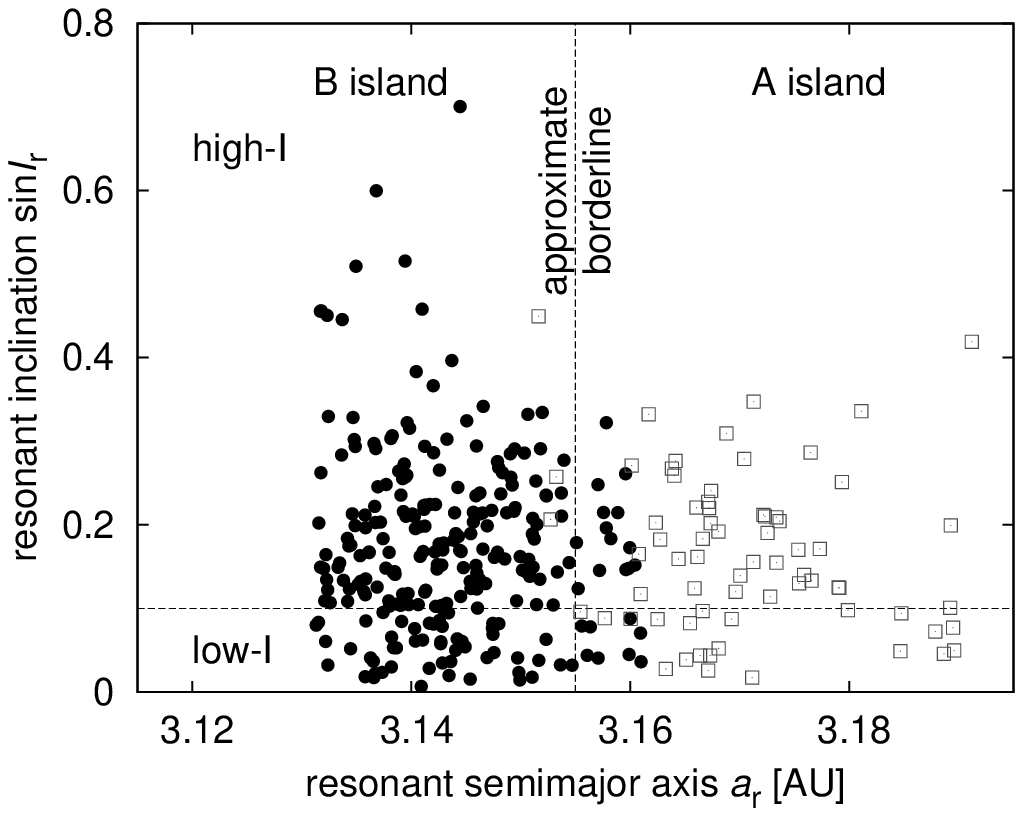}
	\end{tabular}
	\caption{A final result of resonant capture from the outer main belt
		in the fifth giant planet scenario. Top: the resonant semimajor axis $a_{\mathrm{r}}$
		vs the resonant eccentricity $e_{\mathrm{r}}$. The background of the plot
		is the dynamical map of the 2:1 mean-motion resonance with Jupiter
		computed for the post-migration configuration of the giant planets.
		All white symbols indicate orbits of test particles captured
		during the simulation (they correspond to the final state
		in Fig.~\ref{fig:capture_evol}, but only a part of the phase space
		containing the stable islands is plotted). Test particles embedded
		in the dark regions of the dynamical map are candidates for a long-term
		stability and we distinguish them by circles. Other particles
		are marked by crosses.
		Bottom: the resonant semimajor axis $a_{\mathrm{r}}$
		vs sine of the resonant inclination $\sin{I_{\mathrm{r}}}$.
		We plot only the particles captured inside the stable islands.
		Black circles indicate test particles
		captured in island B and gray open squares indicate those
		captured in A island. The horizontal dashed line is plotted
		for reference, because vast majority of the observed B-island Zhongguos reside
		on low inclinations $\sin{I_{\mathrm{r}}}\le0.1$ and all of the
		observed A-island Zhongguos have high inclinations $\sin{I_{\mathrm{r}}}\ge0.1$.
	}
	\label{fig:capture_final}
\end{figure}

The dynamical map for the 2:1 resonance is shown in the Fig.~\ref{fig:capture_final}.
It covers the phase space in the intervals $a\in(3.115,3.195)\,\mathrm{AU}$, $e\in(0.1,0.5)$
and $I\in(0^{\circ},25^{\circ})$, which were
divided into a grid of $40\times40\times5$ boxes (and we finally took an
average over all sections in the inclination).
We set up and followed three test particles per box
for up to $10\,\mathrm{Myr}$.

Comparing the map with Fig.~\ref{fig:map.present}, one can
see very similar structures. The
stable islands A and B are there, separated by the $\nu_{16}$ secular
resonance. The Kozai resonance separatrix can also be easily identified.
A major difference is the shape and size of the A island --
the $\nu_{5}$ resonance is not present in the depicted part of the phase space, which
effectively enlarges the A island with respect to the present state.
The selected candidates for long-lived orbits are also shown in Fig.~\ref{fig:capture_final}.
Note that there are very few A-island bodies captured on highly eccentric
orbits, therefore the aforementioned difference in shape of
island A should not strongly affect the results.
Using the dynamical mapping, we identified $N_{\mathrm{A}}^{\mathrm{synth}}=69$
candidates in island A and $N_{\mathrm{B}}^{\mathrm{synth}}=254$
candidates in island B.

The next step is a rescaling of the initial population
so its particle density would reach realistic values.
We used the observed particle density in the main belt
for bodies with diameters $D\ge5\,\mathrm{km}$ within
the following region: $a\in(2.95,3.21)\,\mathrm{AU}$,
$e\in(0,0.35)$, $I\in(0^{\circ},15^{\circ})$. The intervals
of $e$ and $I$ correspond to those of our synthetic initial population.
On the other hand, the interval of semimajor axes was shifted into the 
region which is currently not depleted by the presence
of the J2/1. We further increased this particle
density by a factor of three \citep{Minton_Malhotra_2010Icar..207..744M}
to account for the dynamical depletion of the main belt 
after the reconfiguration of planets during the last
$\simeq3.85\,\mathrm{Gyr}$. If we increase the particle density
throughout the initial population,
the number of captured long-lived bodies larger than $5\,\mathrm{km}$
would be $N_{\mathrm{A}}^{\mathrm{scaled}}=1552$ and
$N_{\mathrm{B}}^{\mathrm{scaled}}=5857$ in island A and B,
respectively.

Finally, we assume that the dynamical depletion rate due to
chaotic diffusion is the same for
the post-migration islands as for the islands in the observed configuration.
Knowing the approximate number of captured long-lived
asteroids, we employ an exponential decay law and the extremal
values\footnote{These values were derived in Section \ref{sec:islands_decay}
but we use a slightly higher standard deviation (0.05 instead of 0.02).
The reason is that the e-folding times 
were obtained from a model with seven planets and
the Yarkovsky drift included, with long-lived asteroids
of {\em all} sizes. In this section, however, we focus on 
asteroids larger than $5\,\mathrm{km}$ and consequently,
the e-folding times are rather upper limits for us.
We thus artificially increase the original deviation 
to account for this difference.}
of the e-folding times in island A and B 
$\tau_{\mathrm{A}}=\left(0.57\pm0.05\right)\,\mathrm{Gyr}$
and $\tau_{\mathrm{B}}=\left(0.94\pm0.05\right)\,\mathrm{Gyr}$.
We use $t=\left(3.9\pm0.1\right)\,\mathrm{Gyr}$ as a value of the
decay time period.
Combining the results together and identifying the limit values,
our model predicts that we should observe $N_{\mathrm{A}}^{\mathrm{model}}=1$--$3$
long-lived bodies larger than $5\,\mathrm{km}$
in island A and $N_{\mathrm{B}}^{\mathrm{model}}=62$--$121$
asteroids in island B. The observed values are $N_{\mathrm{A}}^{\mathrm{obs}}=2$
and $N_{\mathrm{B}}^{\mathrm{obs}}=71$. The results of our model
and the observed values are in good agreement, which supports
the resonant capture scenario.

Turning our attention to the bottom panel in Fig.~\ref{fig:capture_final},
we can discuss whether the inclination distribution
of captured synthetic bodies
can evolve towards the observed one.
It can be clearly seen that the
number of low-$I$ asteroids in island A is lower than the
number of high-$I$ asteroids. Assuming strong depletion
of the whole island and considering that only $1$--$3$ objects
larger than $5\,\mathrm{km}$ may survive up to the present
according to our model,
the low-$I$ asteroids are more likely to be depleted completely.
That is in agreement with the observations
because the observed asteroids in island A
reside exclusively on highly inclined orbits.

On the other hand, the population captured
in island B stretches almost uniformly over a relatively
large interval of inclinations. Considering the shape
of the real island shown in Fig.~\ref{fig:map.present},
which shrinks with increasing inclination,
the long-term diffusion in the high-$I$ region will
partially deplete this part of the captured population.
But in order to reproduce the observed state in island B,
it is necessary for the source population to contain 
low-$I$ asteroids predominantly.

\subsection{Survival of the primordial population}
\label{sec:survival_primpop}

Let us study hypothetical long-lived primordial orbits and their
survival during the planetary migration. The procedure is similar
to that in the previous section, the major difference is of course
in the initial conditions setup. Here we aim to study only the bodies
which reside inside the J2/1 when the instability simulation begins.
Another important property that should be satisfied is the long-term
stability of the initial orbits -- otherwise one could unintentionally
simulate different effects such as survival of the short-lived
population, etc.

\paragraph*{Initial conditions.}
How to create a set of test particles on the long-lived orbits at the
beginning of the migration scenario? The only viable solution
is to use the dynamical mapping once again. To this effect,
we slightly modify the
standard procedure described in Section \ref{sec:dynmap}. First, the integration
is not done in a stable unvarying configuration of planets. Is is
instead carried out during the first $5\,\mathrm{Myr}$ of the
prescribed migration scenario, starting in the pre-migration configuration
of planets and evolving into a configuration closely preceding the
instability. 

Second, we use only the recorded changes of
the resonant inclination $\delta\sin{I_{\mathrm{r}}}$ when
constructing the map (i.e. we omit $\delta{a_{\mathrm{r}}}$
and $\delta e_{\mathrm{r}}$ when evaluating the metric given
by equation~($\ref{metric}$)).
This is necessary because
as Jupiter migrates inwards, the resonance follows and the 
resonant semimajor axis $a_{\mathrm{r}}$ of bodies residing
inside decreases. This in turn changes the resonant eccentricity
$e_{\mathrm{r}}$ because of the adiabatic invariance of resonant
orbits. These changes are systematic and have different
amplitudes for different orbits. Consequently, they should not be incorporated
when calculating the distance $d$. On the other hand, our
numerical runs suggest that the resonant inclination $I_{\mathrm{r}}$
of test particles does not undergo substantial systematic changes
under the influence of migrating giant planets.
If a large change in inclination is registered, it usually means
that a secular resonance affected the orbit. The resulting
dynamical map thus represents what we need -- the location
of regions crossed by secular resonances and the stable islands
lying in between for the time period before Jupiter's jump.

\begin{figure}
\centering
\includegraphics[width=84mm]{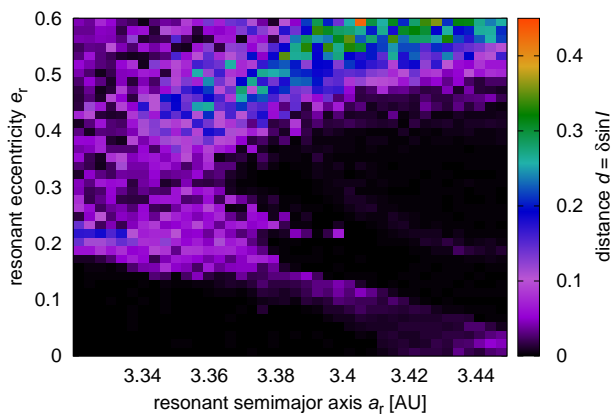} \\
\caption{A dynamical map of the 2:1 mean-motion resonance
	with Jupiter computed during the pre-instability
	evolution of giant planets. The map is displayed
	in the resonant $\left(a_{\mathrm{r}},e_{\mathrm{r}}\right)$ plane
	corresponding to $I_{\mathrm{r}}=2.5^{\circ}$. 
	Note that only the displacement in the resonant 
	inclination $\delta I_{\mathrm{r}}$ is used to 
	represent the dynamical stability.
}
\label{fig:premig.map}
\end{figure}

Our choice of boundaries in the phase space to construct 
the dynamical map was
$a\in(3.32,3.45)\,\mathrm{AU}$ and $e\in(0,0.6)$.
We created $40\times40$ grid in the $(a,e)$ plane and covered
it uniformly with six particles per each cell,
assigning a fixed value of $I=2.5^{\circ}$ to all of them.
As already mentioned, the planetary migration is not able
to strongly change the inclinations, that is why we
map the region of low inclinations where the observed
long-lived population dominates.
The resulting pre-instability dynamical map is displayed
in Fig.~\ref{fig:premig.map}.

\begin{figure*}
\centering
\includegraphics[width=0.82\textwidth]{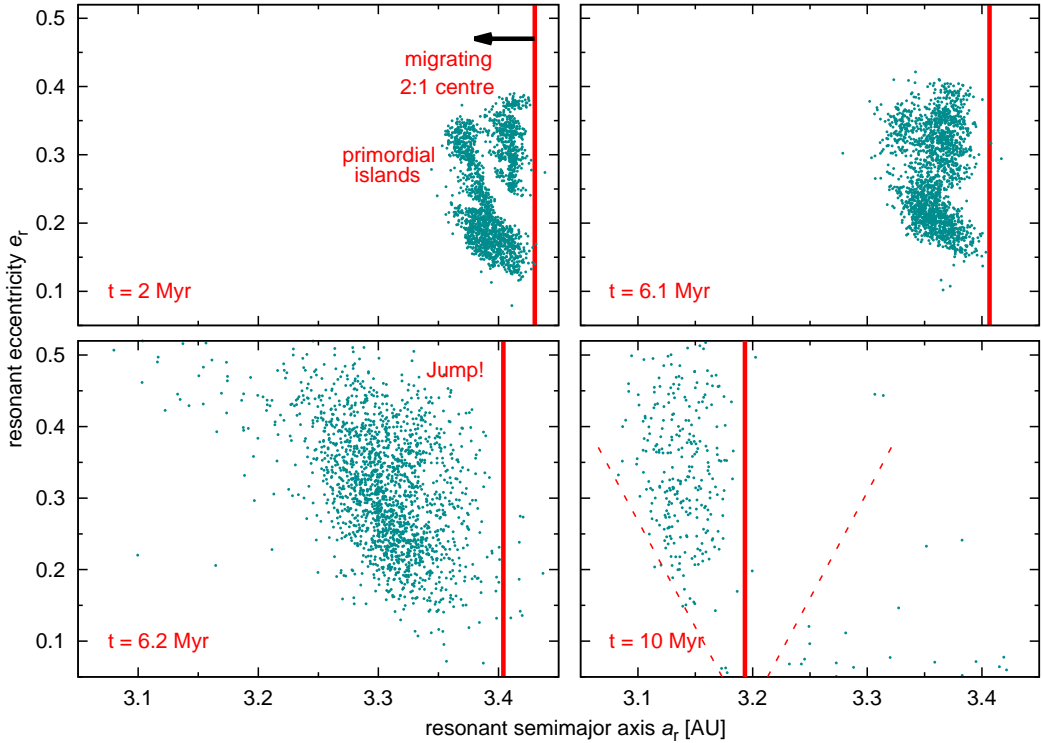} \\
\caption{A simulation of the primordial population survival in the fifth giant planet scenario
	\protect\citep{Nesvorny_Morbidelli_2012AJ....144..117N}.
	The evolution of primordial long-lived orbits in the
	$\left(a_{\mathrm{r}},e_{\mathrm{r}}\right)$ plane is displayed.
	The vertical line indicates the migrating 2:1 resonance.
	The initial population was placed inside the islands identified
	in Fig.~\ref{fig:premig.map}, top left panel represents its orbital distribution
	$2\,\mathrm{Myr}$ after the onset of the migration. The following
	panels show the situation during Jupiter's jump and also the final state.
}
\label{fig:primor_evol}
\end{figure*}

We randomly distributed a group of $2,000$ test particles
over the identified stable islands (see Fig.~\ref{fig:primor_evol})
with low inclinations
$I<5^{\circ}$ to set up a synthetic primordial population
for our simulation. The angular osculating elements were chosen in
such a way
that condition (\ref{reson}) holds and the osculating elements
are therefore identical to the resonant elements at $t=0$.

\paragraph*{Evolution of test particles.}
$2\,\mathrm{Myr}$ after the onset of the migration,
the orbits are not significantly
dispersed which is in agreement with our aim to study only
the dynamically stable resonant bodies.
After $6.1\,\mathrm{Myr}$,
the close encounters of giant planets begin to occur
and start to perturb the islands inside the J2/1. Progression of
Jupiter's jump partially destabilizes the islands and enables significant
number of bodies to leak out and also to populate other regions
of the 2:1 resonance.
At the end of migration, $85$ per cent of the initial population is lost and the rest
is dispersed all over the resonant region, except for the low
eccentricity region. This indicates that the only possible change
in resonant eccentricity during the migration of Jupiter is to increase.

\begin{figure}
\centering
\begin{tabular}{c}
\includegraphics[width=84mm]{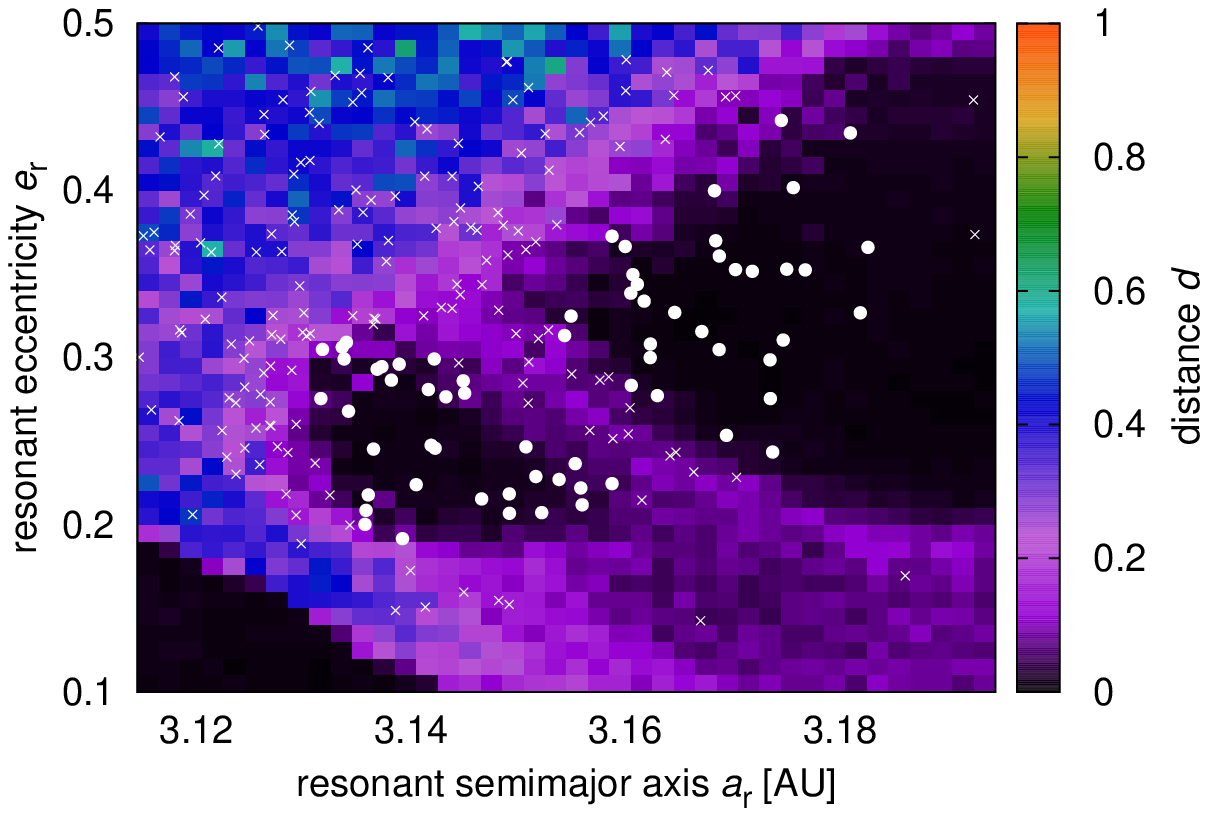} \\
\includegraphics[width=84mm]{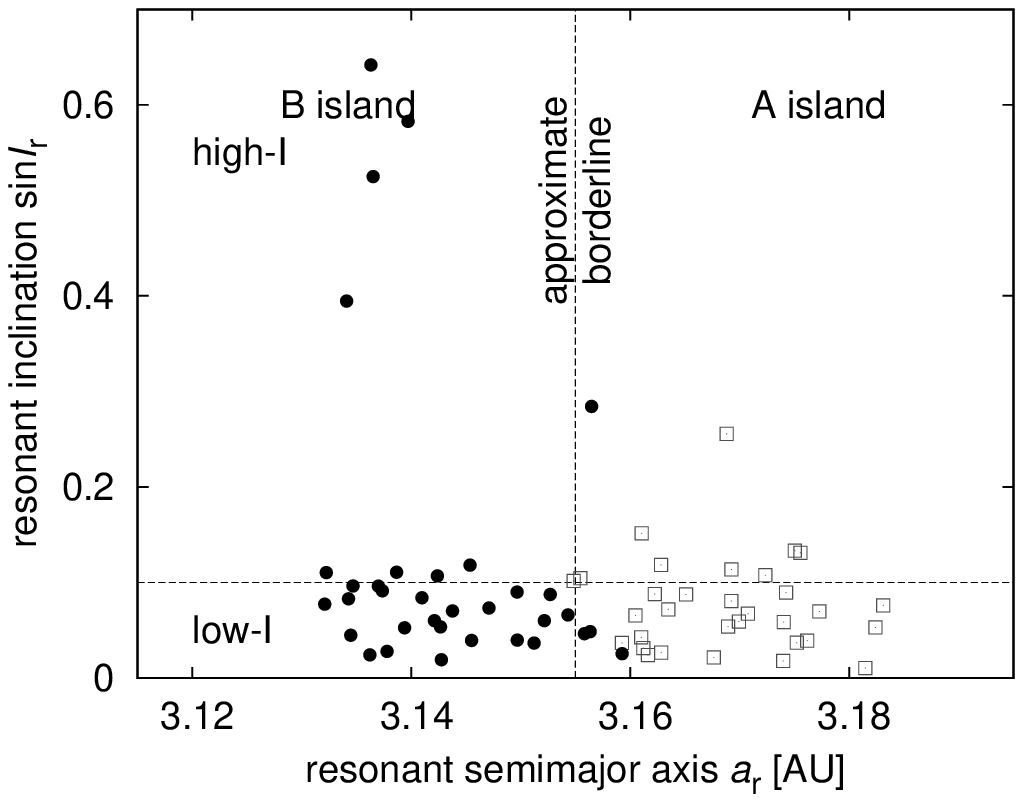}
\end{tabular}
\caption{Top: The same post-migration dynamical map as in Fig.~\ref{fig:capture_final}.
The primordial test particles surviving the jumping-Jupiter instability
are plotted over the map. Circles represent candidates for long-lived orbits
and crosses are the surrounding orbits.
		Bottom: the resonant semimajor axis $a_{\mathrm{r}}$
		vs sine of the resonant inclination $\sin{I_{\mathrm{r}}}$.
		We plot only the particles identified above as long-lived.
		Black circles indicate those
		residing in the island B and gray open squares indicate those
		residing in the A island.
}
\label{fig:primor_final}
\end{figure}

\paragraph*{Surviving population.}
Following our procedure from Section \ref{sec:capture},
we used the same dynamical map
to identify the long-lived asteroids (see
Fig.~\ref{fig:primor_final}) and we applied the rescaling
and the long-term dynamical decay.
Because the particle density in a hypothetical primordial population is
not known, we varied the number of initial synthetic long-lived
particles and calculated the expected number of asteroids
in islands A and B surviving up to the present.

The selected results are listed in Table~\ref{tab:primor.table}
and the comparison is made for asteroids with $D\ge5\,\mathrm{km}$
(we remind the reader that there are $N_{\mathrm{A}}^{\mathrm{obs}}=2$
and $N_{\mathrm{B}}^{\mathrm{obs}}=71$ larger than $5\,\mathrm{km}$
in the observed population). If we assume the particle density
from \cite{Minton_Malhotra_2010Icar..207..744M}, which we also used
in Section~\ref{sec:capture}, the contribution of the primordial
asteroids to the observed population is negligible after long-term
dynamical evolution.

In order to observe a substantial contribution, the particle density in the J2/1
would have to be considerably larger than in the neighbouring main belt
(at least ten times larger).
Such particle density gradient is not very probable, because the
2:1 resonance would have to be dynamically protected against
depleting mechanisms and perturbations, arising e.g. from planetary
embryos \citep*{O'Brien_etal_2007Icar..191..434O} or due to the
compact primordial configuration of planetary orbits
\citep{Masset_Snellgrove_2001MNRAS.320L..55M,Roig_Nesvorny_2014DPS....4640001R}.

\begin{table}
\begin{tabular}{ccc}
\hline
$N_{\mathrm{init}}(D\ge5\,\mathrm{km})$ & $N_{\mathrm{A}}^{\mathrm{model}}$ & $N_{\mathrm{B}}^{\mathrm{model}}$  \\
\hline
2000 &	0 &	0 -- 1 \\
5000 &	0 &	1 -- 2 \\
10000 &	0 &	2 -- 4 \\
100000&	1 -- 3 & 18 -- 35 \\
\hline
\end{tabular}
\caption{The numbers $N_{\mathrm{A}}^{\mathrm{model}}$
and $N_{\mathrm{B}}^{\mathrm{model}}$  of the asteroids
surviving in the islands
A and B up to the present as predicted by our model
of the primordial population survival. The table represents
how the resulting population changes with increasing
number $N_{\mathrm{init}}$ of initial primordial asteroids. The bodies
with $D\ge5\,\mathrm{km}$ are considered. First line roughly corresponds
to the primordial population with particle density of the present
outer main belt, the next two cases approximately consider the particle density
proposed in \protect\cite{Minton_Malhotra_2010Icar..207..744M}. In the last line,
we assume particle density of primordial main belt shortly after its creation,
as suggested
by \protect\cite{Morbidelli_etal_2009Icar..204..558M}; we note that this
case is not very probable.}
\label{tab:primor.table}
\end{table}

We also investigated if A-island orbits can become
B-island and vice versa.
We found that $<1$ per cent of the asteroids initially placed
inside island A survive there and $1$ per cent drift into
B island during the migration. On the other hand, the primordial
population of island B preserves $2$ per cent of the original bodies
and $2$ per cent of them populate island A. Because the primordial
B island is larger than the A island, it harbours more primordial
asteroids under the assumption of homogeneous particle density. The
contribution of island B to the surviving long-lived
population therefore dominates. Also note that after the migration
the ratio $N_{\mathrm{A}}/N_{\mathrm{B}}\simeq1$, thus
the migration itself is not able to create an asymmetric population
out of surviving primordial asteroids. Further orbital
evolution is needed in this case.

The orbital distribution in the
$(a_{\mathrm{r}},I_{\mathrm{r}})$ plane at the end of the planetary
migration indeed resembles the initial interval of inclinations, as both islands
are populated by low-$I$ orbits. To obtain high-$I$ orbits,
which are observed in island A, we would have to assume the presence
of their analogues in the pre-migration islands. But this problem is redundant since
the primordial population probably do not survive up to the present,
as we argued above.

\section{Effects of Jupiter--Saturn great inequality}
\label{sec:GI}
In Section~\ref{sec:effects_of_jj_instability}, we investigated
evolution of the resonant population during a major instability
of the planetary system known as Jupiter's jump. 
In the original migration experiments performed in \cite{Nesvorny_Morbidelli_2012AJ....144..117N},
the violent evolution of planetary orbits was usually followed by
a residual smooth migration during which giant planets
slowly approached their current orbits.
We are in a similar situation,
the planetary configuration at
the end of our instability simulations
slightly differs from the observed one.
In principle, this fact does not affect our ability
to locate and describe the long-lived resonant population of test
particles because we employed fast and efficient method of dynamical
mapping. The natural question then arises,
whether the late migration stages
can invoke perturbations that would affect the
overall stability of the 2:1 resonance significantly.

In particular, we want to check whether the period $P_{\sigma}$
of libration of the resonant asteroids can be comparable with
the Jupiter--Saturn great inequality period $P_{\mathrm{GI}}$
(which is the period of circulation of the
$2\lambda_{\mathrm{J}}-5\lambda_{\mathrm{S}}$ angle)
and what effect would it have on the resonant population
\citep[][see also Section~\ref{sec:intro}]{Ferraz-Mello_etal_1998AJ....116.1491F}.

In order to mimic the smooth late planetary migration
and lead giant planets towards their current orbits,
we used a modified version of the \textsc{swift\_rmvs3}
integrator developed in \cite{Broz_etal_2011MNRAS.414.2716B}.
The integrator introduces an ad hoc dissipation term which
modifies planetary velocity vectors $\bm{v}$ in each time
step $\Delta t$ according to the following relation:
\begin{eqnarray}
	\bm{v}\left(t+\Delta t\right) = \bm{v}\left(t\right)\left[1+\frac{\Delta v}{v}\frac{\Delta t}{\tau_{\mathrm{mig}}}\exp\left(-\frac{t-t_{0}}{\tau_{\mathrm{mig}}}\right)\right] \, ,
	\label{eq:dissipation}
\end{eqnarray}
where $\Delta v = \sqrt{GM/a_{\mathrm{init}}} - \sqrt{GM/a_{\mathrm{fin}}}$
is the total dissipation determined as the difference
of the initial and final mean velocity,
$\tau_{\mathrm{mig}}$ is the migration time-scale,
$t$ is the time variable and $t_{0}$ is an arbitrary
initial time. The eccentricity damping is also included
and can be set independently for each planet by choosing
the damping parameter denoted as~$e_{\mathrm{damp}}$
\citep{Morbidelli_etal_2010AJ....140.1391M}.

To set up the model, we used the final configuration
of giant planets and test particles from our simulations
of resonant capture\footnote{The results of simulations
with primordial asteroids are not considered here, 
as these asteroids probably do not significantly
contribute to the observed population.},
but we only selected
test particles located in broader surroundings of the
stable islands to discard major part of short-lived
asteroids. After this procedure, we launched a set of integrations
with different values of $e_{\mathrm{damp}}$ and different
migration time-scales, bearing in mind that the time-scale
of the original experiments in \cite{Nesvorny_Morbidelli_2012AJ....144..117N}
was $\tau_{\mathrm{mig}}\simeq30\,\mathrm{Myr}$.
When the integrations finished, we selected only the runs
in which giant planets ended up with orbital parameters similar
to the observed ones and we investigated the results at the
time $t_{\mathrm{fin}}$ when $P_{\mathrm{GI}}=880\,\mathrm{yr}$ and the 
ratio of the Saturn's and Jupiter's orbital periods is approximately
$P_{\mathrm{S}}/P_{\mathrm{J}}\simeq2.49$.

\begin{figure}
        \centering
        \includegraphics[width=84mm]{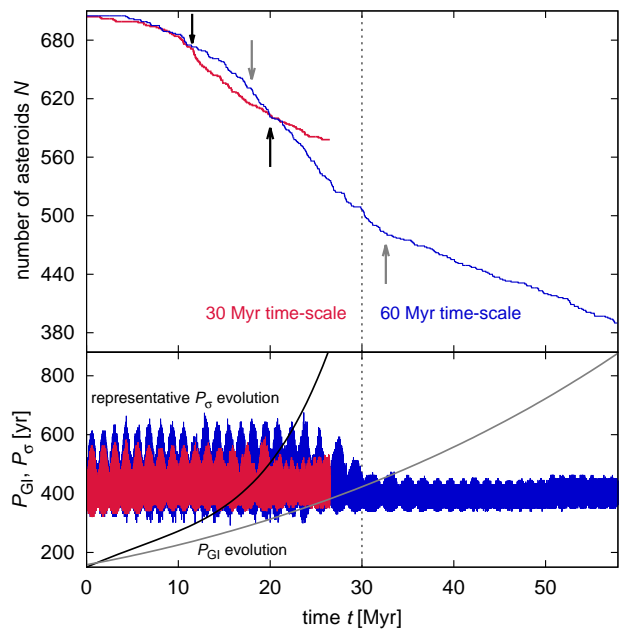}
	\caption{Results of two simulations of smooth
		late migration and its effects on the resonant population.
		Two runs with $t_{\mathrm{fin}}\simeq27\,\mathrm{Myr}$ (red
		and black curves and arrows)
		and $t_{\mathrm{fin}}\simeq58\,\mathrm{Myr}$ (blue and gray curves
		and arrows) are presented
		(migration time-scales $\tau_{\mathrm{mig}}=30\,\mathrm{Myr}$
		and $\tau_{\mathrm{mig}}=60\,\mathrm{Myr}$ were initially chosen but
		the integration is evaluated only until the GI period reaches $880\,\mathrm{yr}$).
                Top: temporal evolution of the number $N$ of test particles.
		Bottom: temporal evolution of the GI period $P_{\mathrm{GI}}$
		and the libration period~$P_{\mathrm{\sigma}}$. The latter is plotted
		for two representative cases. The arrows in the top panel
		approximately mark the time period during which
		$P_{\mathrm{GI}}\simeq P_{\mathrm{\sigma}}$.
                We note that `real' $P_{\mathrm{GI}}$ evolves in an oscillatory manner
		but only with a small amplitude; a polynomial fit of the real
		$P_{\mathrm{GI}}$
		evolution is plotted here for clarity.}
        \label{fig:ntp_gi_sigma}
\end{figure}

In the following, we will discuss results of two runs with
$t_{\mathrm{fin}}\simeq27\,\mathrm{Myr}$ and $t_{\mathrm{fin}}\simeq58\,\mathrm{Myr}$.
Fig.~\ref{fig:ntp_gi_sigma} shows the temporal evolution
of the number $N$ of test particles in these two runs,
the GI period $P_{\mathrm{GI}}$
and the libration period $P_{\sigma}$ of two typical
resonant asteroids. While $P_{\sigma}$ oscillates 
around a nearly constant value $\simeq420\,\mathrm{yr}$,
$P_{\mathrm{GI}}$ initially lies below this value
and rises smoothly as Jupiter and Saturn undergo divergent
migration, and their eccentricities are damped.

The evolution of the $N(t)$ curves does not follow any
simple decay law.  This may serve as a proof that the
level of chaotic diffusion inside the islands indeed
evolves during the smooth late migration.
The steepest slope of the $N(t)$ dependency
is reached during the time interval when
$P_{\mathrm{GI}}\simeq P_{\sigma}$. But one can also notice
that the slope changes are smooth rather than sharp;
the destabilization is not strictly bounded by
the $P_{\mathrm{GI}}\simeq P_{\sigma}$ condition.
This is in agreement with the results of
\cite{Ferraz-Mello_etal_1998AJ....116.1491F}
who argued that even a near-resonant state of
$P_{\mathrm{GI}}$ and $P_{\sigma}$ can enhance the chaotic
diffusion.
However, it is clear that
the number of asteroids surviving the smooth
late migration depends mainly on the length of the time
interval during which $P_{\mathrm{GI}}\simeq P_{\sigma}$;
the population is more depleted in the
run with $t_{\mathrm{fin}}\simeq58\,\mathrm{Myr}$. 

With the aim to evaluate the result of the models with
residual migration included,
we first identified test particles which were located
inside the stable islands at the time~$t_{\mathrm{fin}}$.
At this time, the orbital architecture of giant planets
approximately corresponds to the observed one
and therefore the identification of long-lived asteroids
can be achieved using the {\em current} dynamical map.
Finally, we derived expected numbers
$N_{\mathrm{A}}^{\mathrm{model}}$ and
$N_{\mathrm{B}}^{\mathrm{model}}$  of asteroids that
should be observable in the stable islands
after $4\,\mathrm{Gyr}$ of orbital evolution.
For this purpose, we rescaled the initial population
of test particles and we applied simple exponential
dynamical decay law, exactly as in Section~\ref{sec:capture}.

We calculated following values for asteroids larger than~$5\,\mathrm{km}$:
$N_{\mathrm{A}}^{\mathrm{model}}=0$--$1$ and $N_{\mathrm{B}}^{\mathrm{model}}=40$--$78$
in case of $t_{\mathrm{fin}}\simeq27\,\mathrm{Myr}$
and $N_{\mathrm{A}}^{\mathrm{model}}=0$--$1$ and $N_{\mathrm{B}}^{\mathrm{model}}=22$--$42$
in case of $t_{\mathrm{fin}}\simeq58\,\mathrm{Myr}$
(while the observed values are $N_{\mathrm{A}}^{\mathrm{obs}}=2$ and
$N_{\mathrm{B}}^{\mathrm{obs}}=71$).
We can conclude that the hypothesis of resonant capture
is still valid and explains the existence\footnote{Although the values derived
here do not overlap the observations in case of island A, the
difference is only one asteroid. We think that this discrepancy
is not significant as we are comparing small numbers
and the evolution is definitely stochastic.}
of the long-lived population,
if the giant planets finished their late stage
migration on a time-scale comparable to
$\tau_{\mathrm{mig}}\simeq 30\,\mathrm{Myr}$.
Larger time-scales lead to a slower passage of the GI period
over the values of the J2/1 libration period and this
in turn causes an enhanced depletion of the stable islands.
Even in such a case, a significant part of the observed 
population should originate from the resonant capture.

\section{Collisional models}
\label{sec:collisional_models}

Our results of Section~\ref{sec:effects_of_jj_instability} and \ref{sec:GI}
imply that the long-lived J2/1 population was created by resonant
capture during planetary migration. In this section, we further
develop the framework of this hypothesis. We study 
collisional evolution of the J2/1 in order to confirm whether
the size-frequency distribution of the long-lived resonant asteroids can survive
a time period of $4\,\mathrm{Gyr}$ in a non-stationary state.
We thus have to also account for the epoch of the late heavy bombardment (LHB)
\citep{Gomes_etal_2005Natur.435..466G,Levison_etal_2009Natur.460..364L}
during which the transneptunian disc was destabilized and the flux 
of cometary projectiles through the solar system increased substantially.

In the following collisional models,
three populations
of minor solar-system bodies are included: main-belt asteroids,
transneptunian comets and Zhongguos, i.e. we consider only a
subset of the long-lived population. The reason for this simplification
is that the observed SFDs of long-lived
population and its dynamical subgroups share similar features
(see Fig.~\ref{fig:sfds}),
but Zhongguos have the steepest slope $\gamma=-5.1$. Solving an
inverse problem, we
aim to explain the formation of an initial SFD with
an even steeper slope, which would evolve towards
the observed one due to collisions over the $4\,\mathrm{Gyr}$ timespan.
Because of certain level of parametric freedom in collisional models,
the same process of formation should also apply to Griquas with
shallower observed SFD.

\subsection{Intrinsic probabilities and impact velocities}
\label{sec:probas_impvels}

As the first step in the construction of a collisional model, we compute
the intrinsic probability $P_{\mathrm{i}}$ and the mean impact velocity $V_{i}$
of colliding main-belt
and long-lived resonant asteroids. We adopt the method of
\cite{Bottke_etal_1994Icar..107..255B}
based on a geometrical formalism of orbital encounters introduced by
\cite{Greenberg_1982AJ.....87..184G}. To make our samples of orbits 
large enough, we input all long-lived orbits and the first $50,000$ 
main-belt objects from the AstOrb catalogue.

We split the long-lived population into Zhongguos, Griquas with inclination $I\le8^{\circ}$
and Griquas with inclination $I>8^{\circ}$ and calculate the intrinsic probability $P_{\mathrm{i}}$
and weighted mean impact velocity $V_{\mathrm{i}}$ for them, separately.
Using this separation, we want to check if the intrinsic
probabilities for Griquas may differ from Zhongguos significantly.

The results are summarized in the Table~\ref{tab:pivi}. For reference,
\cite{Dahlgren_1998A&A...336.1056D} computed
$P_{\mathrm{i}}= 3.1\times10^{-18}\,\mathrm{km}^{-2}\mathrm{yr}^{-1}$
and $V_{\mathrm{i}} = 5.28 \, \mathrm{km}\,\mathrm{s}^{-1}$ for collisions
between main-belt asteroids. We can see that both Zhongguos and
low-inclined Griquas have $P_{\mathrm{i}}$ and $V_{\mathrm{i}}$
only slightly higher than these reference values. This is caused
by the moderate values of eccentricities in the J2/1 population.
The orbits then more likely intersect with those in the main belt.
The higher collisional velocity is also plausible because
we are comparing an outer main-belt population
with the rest of the main belt.

On the other hand, Griquas with high
inclinations can avoid intersecting some of main-belt orbits,
thus their intrinsic probability is lower by about a factor of two.
Even at this point, we can conclude that the observed difference
in slopes of SFDs (when Griquas and Zhongguos are separated) 
cannot be explained by the differences in $P_{\mathrm{i}}$ and
$V_{\mathrm{i}}$ because the shallower SFD of Griquas
would require higher $P_{\mathrm{i}}$ in order to collide
more often.

\begin{table}
\begin{tabular}{ccc}
\hline
colliding & $P_{\mathrm{i}}$ & $V_{\mathrm{i}}$  \\
populations & [$10^{-18}\,\mathrm{km}^{-2}\mathrm{yr}^{-1}$]  & [$\mathrm{km}\,\mathrm{s}^{-1}$] \\
\hline
Zhongguos vs MB & $3.82$ & $5.36$ \\
Griquas $\left(I\le8^{\circ}\right)$ vs MB & $3.65$ & $5.52$ \\
Griquas $\left(I>8^{\circ}\right)$ vs MB & $1.81$ & $7.57$ \\
\hline
average & $3.09$ & $6.15$ \\
\hline
\end{tabular}
\caption{The intrinsic probability $P_{\mathrm{i}}$ and
the mean impact velocity $V_{\mathrm{i}}$ for collisions
of main-belt and long-lived J2/1 asteroids.}
\label{tab:pivi}
\end{table}

\begin{figure}
	\centering
	\includegraphics[width=84mm]{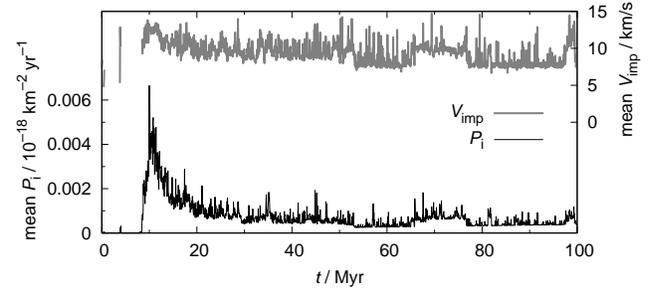}
	\caption{Temporal evolution of the intrinsic probability $P_{\mathrm{i}}(t)$
		(bottom curve)
		and the mean impact velocity $V_{\mathrm{i}}(t)$ (top curve)
		for collisions between transneptunian comets and main-belt
		asteroids, as it was calculated in \protect\cite{Broz_etal_2013A&A...551A.117B}.
		We emphasize that we modify these dependences by numerical factors
		to allow for realistic lifetimes of comets. The values
		$P_{\mathrm{i}}/3$ and $V_{\mathrm{i}}/1.5$ serve as an input
		for our models.}
	\label{fig:pivi_comets}
\end{figure}

\begin{figure}
\centering
\begin{tabular}{cc}
	\includegraphics[width=84mm]{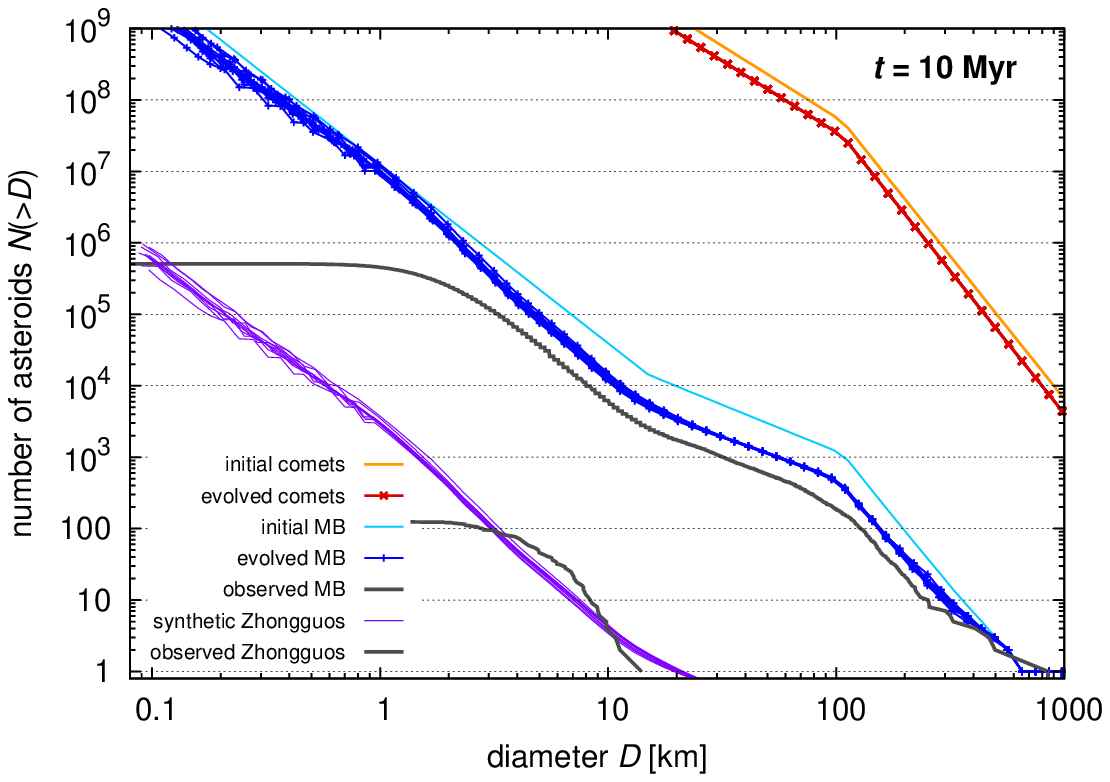} \\
	\includegraphics[width=84mm]{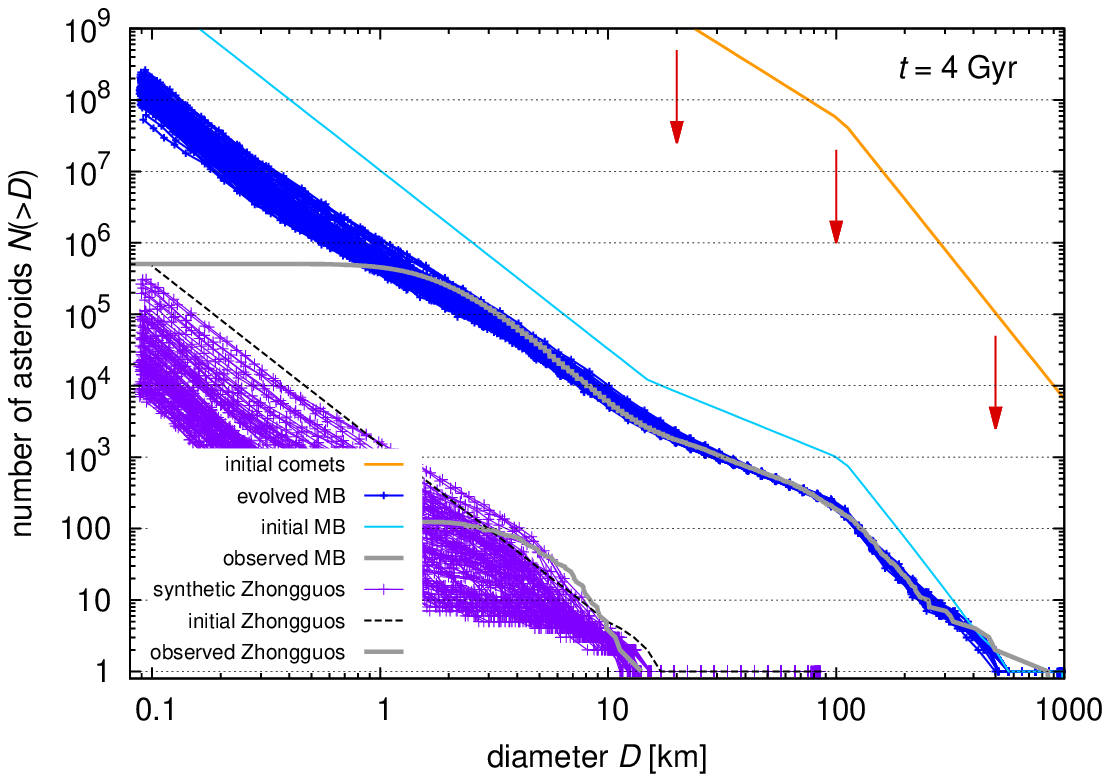} \\
	\includegraphics[width=84mm]{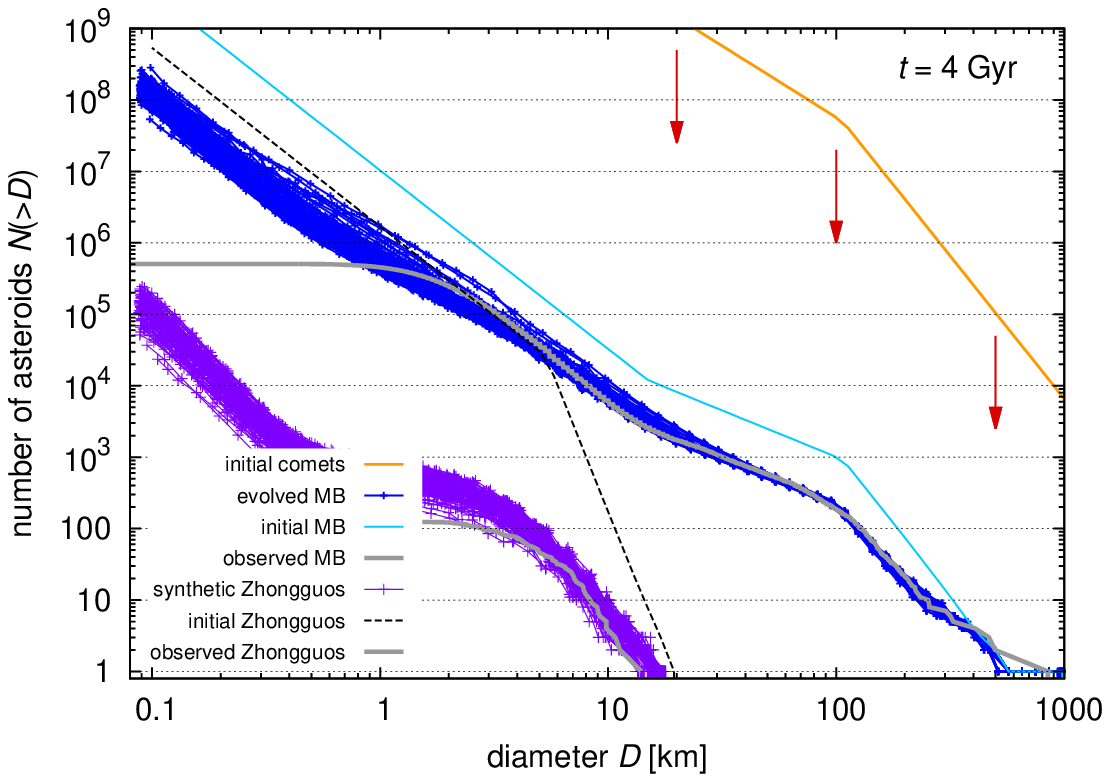} \\
\end{tabular}
\caption{Results of collisional models created with the \textsc{Boulder} code,
	plotted as temporal evolution of the SFDs.
	Description of individual curves is given in the legend of each plot.
	The evolving curves are displayed in the time of the
	simulation $t$, which is also shown.
	Top: synthetic Zhongguos are created by capture
	from the main-belt background population affected by collisions
	with transneptunian comets (see also Section~\ref{subsec:capture_from_MB}).
	Middle: the situation from top left panel is reproduced,
	but a single $D\simeq100\,\mathrm{km}$ asteroid is added to Zhongguos
	and we study the possibility of its catastrophic breakup over $4\,\mathrm{Gyr}$
	(see also Section~\ref{subsec:disruption_PB_inside}).
	Bottom: we demonstrate what parameters of the initial SFD
	of Zhongguos are needed in order to evolve it towards the observed
	one due to collisions (see Section~\ref{subsec:cap_from_family}
	for discussion).
	}
\label{fig:col_models}
\end{figure}

\newpage
\subsection{Capture from the main-belt population}
\label{subsec:capture_from_MB}

Let us ask a question whether the J2/1 SFD
can originate from the main-belt SFD.
We assume that the capture of the resonant population
is {\em not} size dependent and
thus the captured SFD resembles that of the
main belt at the time of planetary migration, scaled down by a
numerical factor. Our arbitrary
choice of this factor is such that the number of the
largest captured bodies is approximately the same as
we observe in the J2/1 population. This allows us
to immediately compare the slope of captured and observed
SFD of Zhongguos.

The main belt itself was likely more
populous $4\,\mathrm{Gyr}$ ago
(by a factor of three according to \cite{Minton_Malhotra_2010Icar..207..744M})
but has remained in a near-stationary collisional regime ever since
\citep{Bottke_etal_2005Icar..175..111B}. The only possible stage of evolution,
which could have temporarily increased the slope of the main belt SFD,
was the late heavy bombardment. We therefore investigate this early
period of collisions between the main belt and transneptunian comets.
We focus on diameters $D<25\,\mathrm{km}$ to see whether
the steepness in this interval can be instantaneously increased -- if so,
a population captured from such SFD would adopt that slope.

We use the \textsc{Boulder} code \citep{Morbidelli_etal_2009Icar..204..558M}
to construct a suitable collisional model. 
We setup initial SFDs by a piecewise power law 
function which is characterized by three differential slope
indices $q_{\mathrm{a}}$, $q_{\mathrm{b}}$ and $q_{\mathrm{c}}$
in the intervals of diameters $D>D_{1}$, $\left(D_{2},D_{1}\right)$ and
$D<D_{2}$, respectively. The SFD is normalized by setting
the number $N_{\mathrm{norm}}$ of bodies larger than $D_{1}$.
The following values are used in case of the main-belt SFD:
$D_{1}=100\,\mathrm{km}$, $D_{2}=14\,\mathrm{km}$, $q_{\mathrm{a}}=-5.0$,
$q_{\mathrm{b}}=-2.3$, $q_{\mathrm{c}}=-3.5$, $N_{\mathrm{norm}}=1110$.
A few $D\sim1000\,\mathrm{km}$ asteroids are added in order
to properly reproduce the present state.
Using the same assumptions as \cite{Broz_etal_2013A&A...551A.117B}, the cometary disc
is characterized as:
$D_{1}=100\,\mathrm{km}$, $q_{\mathrm{a}}=-5.0$,
$q_{\mathrm{b}}=q_{\mathrm{c}}=-3.0$, $N_{\mathrm{norm}}=5\times10^{7}$.

Because the evolution of the transneptunian
disc is dominated by its fast dynamical dispersion, the intrinsic
probability $P_{\mathrm{i}}(t)$ and mean impact velocity $V_{\mathrm{i}}(t)$
of collisions with comets are time-dependent quantities.
The temporal evolution of $P_{\mathrm{i}}(t)$
and $V_{\mathrm{i}}(t)$, which was derived in \cite{Broz_etal_2013A&A...551A.117B},
is shown in Fig.~\ref{fig:pivi_comets}. \cite{Broz_etal_2013A&A...551A.117B}
also argued that it is necessary to modify this dependence
in order to mimic the effects of spontaneous cometary breakups
due to various physical processes. Regarding these effects, they derived quantities
$\tilde{P}_{\mathrm{i}}(t)=P_{\mathrm{i}}(t)/3$ and $\tilde{V}_{\mathrm{i}}(t)=V_{\mathrm{i}}(t)/1.5$
as a feasible modification.

The specific energy $Q_{\mathrm{D}}^{\star}$ required to
disperse $50$ per cent of the shattered target is described by the polynomial scaling law
\citep{Benz_Asphaug_1999Icar..142....5B}
\begin{eqnarray}
	Q_{\mathrm{D}}^{\star}\left(r\right)=\frac{1}{q_{\mathrm{fact}}}\left(Q_{0}r^{a}+B\rho r^{b}\right) \, ,
	\label{eq:scaling_law}
\end{eqnarray}
where $r$ denotes the target's radius in $\mathrm{cm}$.
Material parameters $\rho$, $Q_{0}$, $a$, $B$, $b$ and $q_{\mathrm{fact}}$
for basaltic rock (used for asteroids) and water ice (used for comets)
are listed in Table~\ref{tab:scaling_params}.

\begin{table}
	\centering
	\begin{tabular}{cccccc}
		\hline
		$\rho$ & $Q_{0}$ & $a$ & $B$ & $b$ & $q_{\mathrm{fact}}$ \\
		\lbrack$\mathrm{g}\,\mathrm{cm}^{-3}$\rbrack & \lbrack$\mathrm{erg}\,\mathrm{g}^{-1}$\rbrack & & \lbrack$\mathrm{erg}\,\mathrm{g}^{-1}$\rbrack & & \\
		\hline
		\multicolumn{2}{l}{Basalt:} & & & & \\
		3.0 & $7\times10^{7}$ & $-0.45$ & $2.1$ & $1.19$ & $1.0$ \\
		\multicolumn{2}{l}{Water ice:} & & & & \\
		1.0 & $1.6\times10^{7}$ & $-0.39$ & $1.2$ & $1.26$ & $3.0$ \\
		\hline
	\end{tabular}
	\caption{Material parameters $\rho$, $Q_{0}$, $a$, $B$, $b$ and $q_{\mathrm{fact}}$
		of the polynomial scaling law (see equation~(\ref{eq:scaling_law}))
	adopted from \protect\cite{Benz_Asphaug_1999Icar..142....5B}. The first line is used in case
	of main-belt or resonant asteroids and the second line is used for comets.}
	\label{tab:scaling_params}
\end{table}

We simulated the collisional evolution for a time period of $10\,\mathrm{Myr}$
in which $P_{\mathrm{i}}\left(t\right)$ reaches its maximum (see
Fig.~\ref{fig:pivi_comets} and top panel of Fig.~\ref{fig:col_models}).
The flux of cometary projectiles induced by ongoing planetary migration reaches
its climax and strongly affects the main-belt SFD.
We performed 100 numerical realisations of the model with different
seeds of the built-in random-number generator to account for stochasticity
of the evolution and possible low-probability breakups. At the end of the
simulation, we assume the resonant capture takes place with
the efficiency factor $10^{-4}$. The steep part of the captured
SFD can be characterized by the slope $\gamma = -3.0$. Although
the region of the $\mathrm{km}$-sized
bodies is steeper with respect to the initial state of the main-belt SFD,
it is not steep enough to reach the slope of the SFD of observed Zhongguos.
We thus consider this scenario unlikely.

\subsection{Catastrophic disruption of a captured parent body}
\label{subsec:disruption_PB_inside}
From the result of the previous section, it is obvious that a different
explanation of the steep initial SFD is needed.
Here we test the possibility that a large parent body
was captured inside the 2:1 resonance by chance, it was
subsequently disrupted and its fragments formed a family.

We first use a stationary model to investigate the probability of
such a catastrophic breakup event. As a prerequisite, we estimated
a lower limit of the parent body size $D_{\mathrm{PB}}$.
To achieve that, we searched the outcomes of
hydrodynamic disruption models in
\cite{Durda_etal_2007Icar..186..498D} and \cite{Benavidez_etal_2012Icar..219...57B}
for SFDs which have approximately the same slope as
the observed SFD of Zhongguos. We rescaled the sizes
of synthetic asteroids in these datasets
so that the diameter of the largest remnant
would correspond approximately to the size
of (3789) Zhongguo. We selected all reasonable fits
and derived range of admissible values
$D_{\mathrm{PB}}\in\left(50,120\right)\,\mathrm{km}$
with the median value $\tilde{D}_{\mathrm{PB}}=70\,\mathrm{km}$.

We then use the following relation expressing the number $N_{\mathrm{col}}$
of catastrophic disruptions of parent bodies with the diameter
$D_{\mathrm{PB}}$ due to collisions with a population of projectiles:
\begin{eqnarray}
N_{\mathrm{col}} = P_{\mathrm{i}}N_{\mathrm{PB}}N_{\mathrm{project}}\frac{D_{\mathrm{PB}}^{2}}{4}\Delta t  \, ,
\end{eqnarray}
where $P_{\mathrm{i}}$ denotes the intrinsic probability, $N_{\mathrm{PB}}$
is the number of parent bodies, $N_{\mathrm{project}}$ is the number
of projectiles capable of disrupting the parent body and
$\Delta t$ is the considered timespan.

Since there are {\em no} large bodies observed in the current J2/1 population,
we assume $N_{\mathrm{PB}}=1$. To provide the lower limit on
the diameter of the projectiles $d_{\mathrm{disrupt}}$, we use the model of
\cite{Bottke_etal_2005Icar..179...63B}
in combination with the scaling law given by equation~(\ref{eq:scaling_law})
\citep{Benz_Asphaug_1999Icar..142....5B}.
We then take $N_{\mathrm{project}}$ for different values of $d_{\mathrm{disrupt}}$
as the number of main-belt asteroids with diameters $D\ge d_{\mathrm{disrupt}}$.
Several resulting values of $N_{\mathrm{col}}$ over
$\Delta t=4\,\mathrm{Gyr}$ timespan are given
in Table~\ref{ncol}.
It turns out that the probability
of a catastrophic disruption is $\le18$ per cent when considering
a single target only.

Note that a more realistic case should take into consideration
that there is no observable evidence of a collisional cluster,
thus the hypothetical breakup event
must have occurred more than $1\,\mathrm{Gyr}$ ago
\citep{Broz_etal_2005MNRAS.359.1437B}.
Of course, one should also account for 
the influence of cometary flux during LHB which can temporarily increase
the rate of collisions but also tends to speed up the 
evolution of SFDs.

\begin{table}
\begin{tabular}{cccc}
\hline
$D_{\mathrm{PB}}$ & $d_{\mathrm{disrupt}}$ & $N_{\mathrm{project}}$ & $N_{\mathrm{col}}$ \\
$\lbrack\mathrm{km}\rbrack$ & $\lbrack\mathrm{km}\rbrack$ &   & \\
\hline
50 & 6 & 25821 & 0.18 \\
70 & 9 & 7837 & 0.10 \\
100 & 15 & 2589 & 0.07 \\
120 & 19 & 1777 & 0.07 \\
\hline
\end{tabular}
\caption{The results of a stationary collisional model:
        the number $N_{\mathrm{col}}$ of
        catastrophic breakups of a parent body with the diameter 
	$D_{\mathrm{PB}}$ over $4\,\mathrm{Gyr}$ timespan.
	We also list the minimal size $d_{\mathrm{disrupt}}$
	and the number $N_{\mathrm{project}}$ of suitable projectiles.}
\label{ncol}
\end{table}

We therefore test the possibility of a catastrophic breakup
once again in the more sophisticated
framework of the \textsc{Boulder} code and we also check
the influence of comets.
The setup for main-belt asteroids and comets is the same
as in Section \ref{subsec:capture_from_MB}, as well as the SFD
of synthetic Zhongguos, which is only modified by adding a 
single $D\simeq100\,\mathrm{km}$ asteroid. The intrinsic probability
and the mean impact velocity
for MB vs. Zhongguos collisions are taken as the average of
values from Table~\ref{tab:pivi}. We simulated $4\,\mathrm{Gyr}$
of collisional evolution.

The middle panel of Fig.~\ref{fig:col_models} shows
a set of 100 realisations of our model. At the beginning
of the simulations, the synthetic
population of Zhongguos is strongly affected by the cometary flux,
which mostly causes cratering of the large body in the resonance.
Moreover, a family-forming event takes place in a few runs. Simultaneously,
the comets speed up the collisional evolution in and below the
region of mid sized asteroids. As a result, families
inside the J2/1 are comminutioned too fast and the steepness of their
SFDs drops below the observed one.

When studying the family-forming events after the LHB,
we focused only on cases with the size of the largest 
fragment (or remnant) $D_{\mathrm{LF}}\ge10\,\mathrm{km}$
to obtain a body similar to (3789) Zhongguo.
In three such cases, which occurred at the time of simulation
$t=2.1$, $2.8$ and $3.1\,\mathrm{Gyr}$, a parent body
with the diameter $D_{\mathrm{PB}}\simeq85\,\mathrm{km}$ was
shattered into a family with the largest fragment having
$D_{\mathrm{LF}}\simeq18\,\mathrm{km}$
and the largest remnant having $D_{\mathrm{LR}}\simeq60\,\mathrm{km}$.
Corresponding evolved SFDs match\footnote{In fact, the SFDs
resulting from our simulations lie slightly below the observed case
but this discrepancy can be easily removed assuming a bit larger
parent body.} the observed one,
except for the presence of the parent-body remnant.

We can conclude that a family-forming event inside the
2:1 resonance is a very unlikely process because
a complex set of constraints has to be satisfied:
A major collision would have to occur despite its low probability,
which is typically $10$ per cent, as given by the stationary model,
or $3$ per cent, as derived from the more sophisticated model.
The created SFD would have to be very steep, preferably 
without presence of a parent-body remnant. If the remnant
was present, one would have to rely on its subsequent
elimination due to dynamical depletion because there is
no large asteroid observed in the J2/1. 
Finally, the breakup would have to take place after the LHB,
but early enough to allow for dispersion of the collisional cluster.

\subsection{Capture from a family}
\label{subsec:cap_from_family}

\begin{figure}
\centering
\begin{tabular}{c}
	\includegraphics[width=84mm]{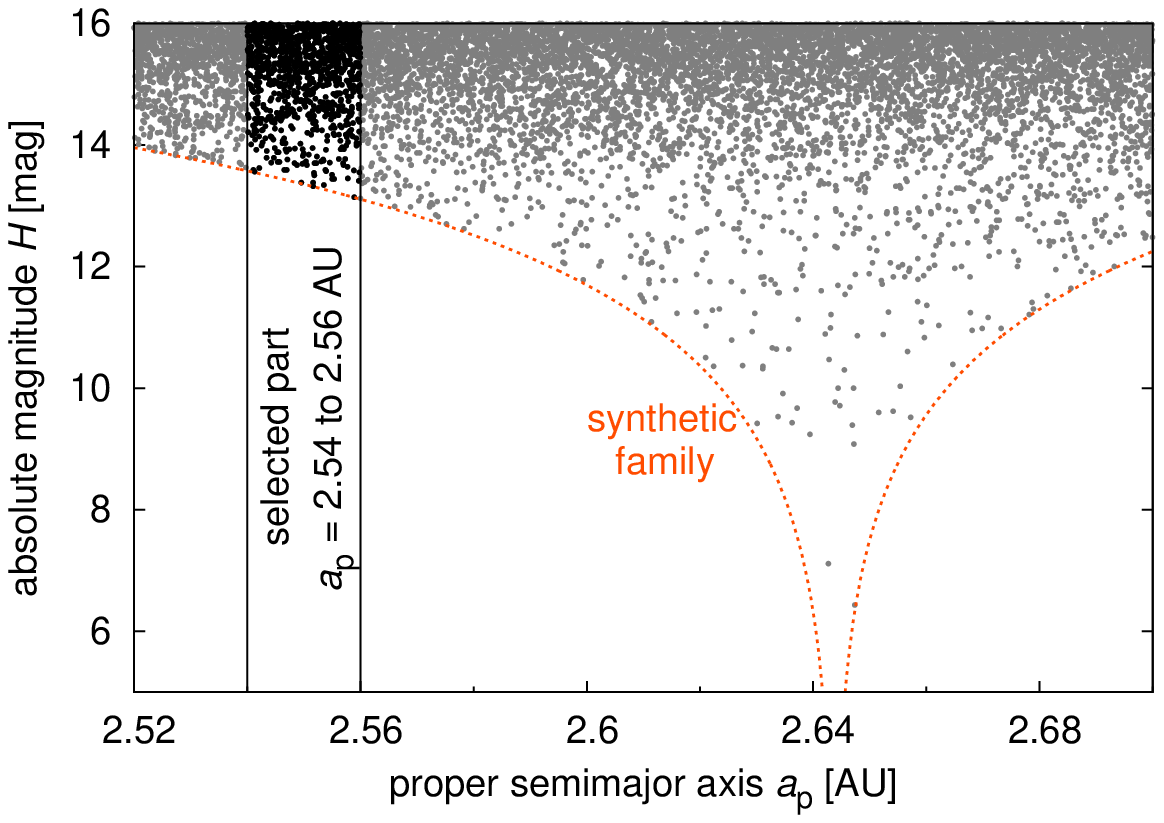} \\
	\includegraphics[width=84mm]{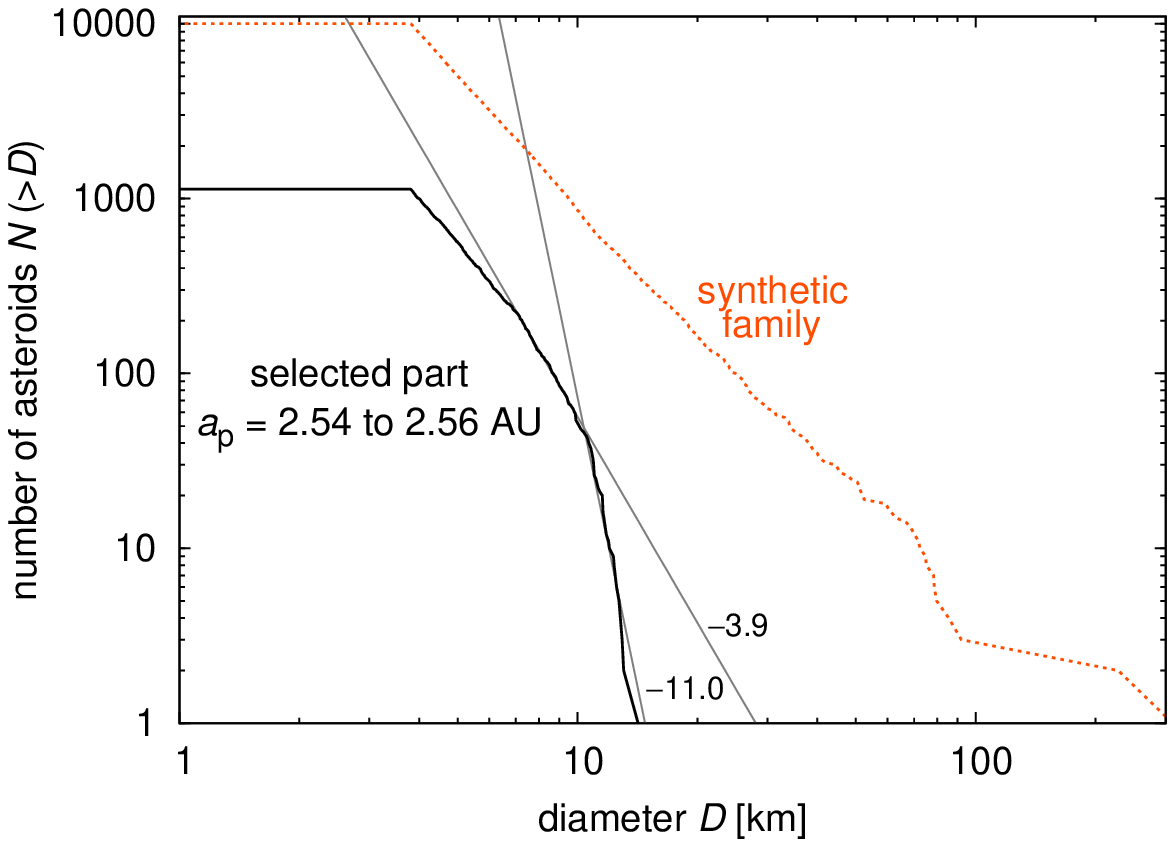} \\
\end{tabular}
	\caption{Dependence of the absolute magnitude $H$
		on the proper semi-major axis $a_{\mathrm{p}}$
		for a synthetic family (top panel).
		The family is used in
		a simple Monte-Carlo test, in which we study
		how the shape of the SFD changes
		if we select only a part of the family.
		An example, for the part of the family
		bordered by the vertical lines, is given
		in the bottom panel.
		We compare cumulative size-frequency distributions
		of the entire synthetic
		family (dashed curve) and of its part (solid curve).
		The steep segments are approximated with a
		power law function and obtained slope indices are shown
		for reference.
		}
	\label{fig:cap_from_family}
\end{figure}

Previous collisional models imply that it is difficult to
explain the steep SFD of Zhongguos
by standard processes (such as direct capture from the background main-belt
population or family-forming event)
that could have occurred during resonant capture or later on.
We thus employ a sort of `reversed' method in this section. We first
try to discover
what parameters of the initial SFD are needed in order to
obtain the observed one after $4\,\mathrm{Gyr}$ of
collisional evolution. Then we discuss possible process which
could have led to formation of such an SFD.

Using trial and error,
we established the following initial SFD as an appropriate one:
$D_{1}=10\,\mathrm{km}$, $D_{2}=5\,\mathrm{km}$, $q_{\mathrm{a}}=-6.6$,
$q_{\mathrm{b}}=-6.6$, $q_{\mathrm{c}}=-3.5$, $N_{\mathrm{norm}}=12$.
The change of slope at $D_{2}$ is necessary to reasonably
satisfy the mass conservation law.
The main-belt and cometary SFDs are the same as in the previous models.

The result of 100 runs of the collisional code is displayed in 
bottom panel of Fig.~\ref{fig:col_models}. The evolved
SFD of the resonant population corresponds to the observed one
very well. The steep parts have the same slope, the only difference
can be seen in the region of small bodies where the synthetic
SFD is abundant. But one has to realize that the observed SFD is
probably biased in this region.

The mechanism responsible for creation of the
high initial slope may be the following (and
it is related to the effect described in 
\cite*{Carruba_etal_2014ApJ...792...46C}).
Let us assume resonant capture from a hypothetical
outer main-belt family instead of capture from 
the background main-belt population. The stable islands 
are able to capture only a fraction
of this family because jump of the resonance
is thought to be fast (it does not sweep through the resonance),
the capture efficiency is limited (see Section~\ref{sec:capture}),
and also the width of the islands in semimajor
axis is relatively small (see e.g. Fig.~\ref{fig:map.present}).
If the captured part of the family is located farther
away from the original position of the parent body,
its SFD will be very steep because
smaller fragments have higher ejection velocities and
therefore land at orbits that are more distant from the
parent body.

We employed a simple Monte-Carlo test to investigate
the described possibility.
We generated a uniform distribution of $10,000$ test particles
satisfying the bounds
$2.52\le a \le 2.70 \,\mathrm{AU}$ and
$\log\left({\left|a-a_{\mathrm{c}}\right|/2\times10^{-4}}\right)/0.2\le H \le 16$
in the $(a,H)$ plane where $a_{\mathrm{c}}\simeq2.64\,\mathrm{AU}$ is
value of the central semimajor axis and $H$ is the absolute magnitude.
Finally, we assigned
diameters $D$ to the test particles on the basis of the $H$
distribution, assuming single
value of the geometric albedo $p_{\mathrm{V}}=0.05$ throughout
our sample. This way we created a synthetic family
(similar e.g. to Eunomia family).
Then we randomly moved a $\Delta a=0.02\,\mathrm{AU}$ window
in the semimajor axis over the collisional cluster and we 
monitored the SFDs in the successively selected regions.

An example is given in Fig.~\ref{fig:cap_from_family}.
Obviously, extremely high slope can be reached by this process.
The resulting slope is well above the estimated lower limit
which is needed for the initial SFD of Zhongguos.
Note that this mechanism does not require the family
{\em itself} to have steep SFD. The steepness is achieved
afterwards by selective resonant capture from an appropriate region.

Finally, we remark that an interesting link can be found between
the hypothesis of capture from a family and our dynamical
simulations. In Section~\ref{sec:capture},
we argued that a source population for resonant capture
should contain greater
number of asteroids with low inclination in order to explain
the observed high concentration of low-$I$ B-island bodies.
Thus we can conclude that if the hypothetical outer
main-belt family was indeed captured, it probably must have been
located on low inclinations.


\section{Conclusions}
\label{sec:conclusions}

Let us briefly review the content of this paper.
We updated the population of asteroids residing in the 2:1
mean-motion resonance with Jupiter using recent observational data.
The new list of resonant objects now contains 370 bodies,
which can be divided on the basis of their mean dynamical
lifetime to 140 short-lived and 230 long-lived asteroids.
Our revision of physical properties of the resonant population
is generally in agreement with the conclusions of previous studies;
new result is our estimate of the mean albedo $\bar{p}_{\mathrm{V}}=(0.08\pm0.03)$,
based on the data from the WISE database.

The long-term dynamics of the two quasi-stable islands A
and B was studied, using the results of Skoulidou et al. (in preparation)
who took into account also the perturbations induced by the
terrestrial planets and the semimajor axis drift caused
by the Yarkovsky effect. The population in the islands
decays exponentially, but the escape rate is significantly
faster in island A, the e-folding times being
$0.57\,\mathrm{Gyr}$ for island A and $0.94\,\mathrm{Gyr}$
for island B. Hence, the dynamical evolution on $\mathrm{Gyr}$-long
time-scales results into differential depletion of the islands,
which is certainly one of the reasons for the observed
asymmetric A/B population ratio.

The primary goal of this paper was to explain the
formation of the long-lived population,
satisfying the observational constraints.
We tested two hypotheses:
\begin{enumerate}
	\item capture from the outer main belt,
	\item survival of primordial long-lived resonant asteroids,
\end{enumerate}
both in the framework
of the five giant planets migration scenario \citep{Nesvorny_Morbidelli_2012AJ....144..117N}
with a jumping-Jupiter instability.

Our simulations of the instability imply that both processes could have been
at work. The capture itself, together with subsequent differential
depletion due to long-term dynamical evolution,
can explain the observed population in both islands,
if we assume that the number of asteroids in the main belt
$4\,\mathrm{Gyr}$ ago was {\em three times larger} than the current one,
as suggested by \cite{Minton_Malhotra_2010Icar..207..744M}.
Namely, the numbers of asteroids with $D\ge5\,\mathrm{km}$,
which should survive to this day, are
$N_{\mathrm{A}}^{\mathrm{model}}=1$--$3$ and 
$N_{\mathrm{B}}^{\mathrm{model}}=62$--$121$, as
predicted by our model.
For comparison, the observed numbers are $N_{\mathrm{A}}^{\mathrm{obs}}=2$
and $N_{\mathrm{B}}^{\mathrm{obs}}=71$.

We also modeled the late residual planetary migration
to check on the effect of possible resonance between the 
libration period in the 2:1 commensurability and 
the raising period of Jupiter--Saturn great inequality.
The impact of this effect on the long-lived population
depends on the time-scale of the late migration.
If the time-scale was $\simeq30\,\mathrm{Myr}$
(which corresponds with the scenarios in
\cite{Nesvorny_Morbidelli_2012AJ....144..117N}),
the destabilization of the islands would be weak enough
for the captured population to survive in a state
similar to the observations. More specifically,
if we account for the effect of great inequality
in our model, the resulting values are
$N_{\mathrm{A}}^{\mathrm{model}}=0$--$1$ and
$N_{\mathrm{B}}^{\mathrm{model}}=40$--$78$.
If the time-scale was $\simeq60\,\mathrm{Myr}$,
the chaos in the stable islands would be intensified
for longer period of time and the depletion of the
captured population would be more significant, leading
to $N_{\mathrm{A}}^{\mathrm{model}}=0$--$1$ and
$N_{\mathrm{B}}^{\mathrm{model}}=22$--$42$.

In case of primordial resonant objects, we discovered
that approximately $1$ per cent of A-island and $4$ per cent
of B-island asteroids survive
Jupiter's jump on long-lived orbits. However,
we argued that the primordial asteroids probably 
do not contribute to the observed population at all.
The reason is that the primordial population
would have to exhibit particle density larger 
than the one proposed by \cite{Minton_Malhotra_2010Icar..207..744M} by a factor
of ten, in order to survive $4\,\mathrm{Gyr}$ of post-migration dynamical evolution.
Such particle density would create a gradient with respect to the neighbouring main belt
which we assume dubious.

Finally, by creating several collisional models, we
demonstrated that the observed steep SFD
of the long-lived asteroids cannot be
explained by capture
from the background main-belt SFD, affected by the late heavy
bombardment. We also proved that a family-forming catastrophic
disruption inside the J2/1 is very unlikely.

Our main conclusion is that the long-lived J2/1 population
was probably formed by capture from a hypothetical
outer main-belt family during Jupiter's jump.
If this is the case, then the long-lived asteroids
in the 2:1 resonance with Jupiter represent
the oldest identifiable remnants of a main-belt asteroidal
family.

There are several improvements that are needed to
conclude the debate on the long-lived population in
the 2:1 resonance and its origin. For example, it is appropriate
to assess the possibility of the Themis family formation event
as a contributor to the resonant population.
Although we did not study this
hypothesis in detail, we summarized several results of our 
preliminary tests in Appendix~\ref{sec:themis}. Our simulations
suggest that this possibility is not viable and this in turn
supports the hypothesis of resonant capture.

Other improvements of our work might include a
development of a self-consistent model of both dynamical
and collisional evolution of long-lived asteroids.
Finally, we suggest to study the hypothetical
primordial population in the epoch {\em before} the
jumping-Jupiter instability in order to properly estimate primordial
particle density inside the stable islands.


\section*{Acknowledgements}

The work of OC and MB has been supported by the Grant Agency
of the Czech Republic (grant no.\ 13-01308S) and by
Charles University in Prague (project GA UK no.~1062214; project SVV-260089).
The work of DN was supported by NASA's Solar System Workings program.
We thank an anonymous reviewer for valuable comments, which improved
the final version of the paper.



\bibliographystyle{mn2e}
\bibliography{references}


\appendix

\section[]{Contribution of the Themis family ejecta to the long-lived population}
\label{sec:themis}

We shall briefly discuss a possible role of neighbouring Themis family
in creation of the long-lived resonant population. Although 
\cite{Broz_etal_2005MNRAS.359.1437B} demonstrated that the fragments
from Themis family cannot be transported to the stable islands by the Yarkovsky
semimajor axis drift, a possibility of direct injection during the Themis
family formation event is still an open question. 
During the formation, fragments ejected at high velocities
($v_{\rm ej} \simeq 400\,\mathrm{m}\,\mathrm{s}^{-1}$, or more) may fall directly
in the J2/1 resonance and thus contribute somehow to the stable population.
There are several caveats, however.

(i)~One has to assume that a substantial part of fragments ($\simeq 10$ per cent)
have very large ejection velocities with respect to (24)~Themis,
or the respective parent body with $D_{\rm PB} \simeq 400\,{\rm km}$.
We tried to use an `extreme' size-independent velocity field
prescribed by \cite*{Farinella_etal_1993Icar..101..174F} relation,
with parameters $v_{\rm esc} = 170\,\mathrm{m}\,\mathrm{s}^{-1}$ and $\alpha = 3.25$.
Then up to $N(D > 5\,{\rm km}) = 10$--$30$ fragments
land within island B, according to our tests. This number should be
further decreased by a subsequent long-term orbital evolution,
as Themis family is $(2.5\pm 1.0)\,{\rm Gyr}$ old \citep{Broz_etal_2013A&A...551A.117B}.
Let us also note that the collisional evolution of resonant bodies
was accounted for automatically, as we did this simulation with
the currently observed size-frequency distribution of Themis family.

\begin{figure}
	\centering
        \includegraphics[width=84mm]{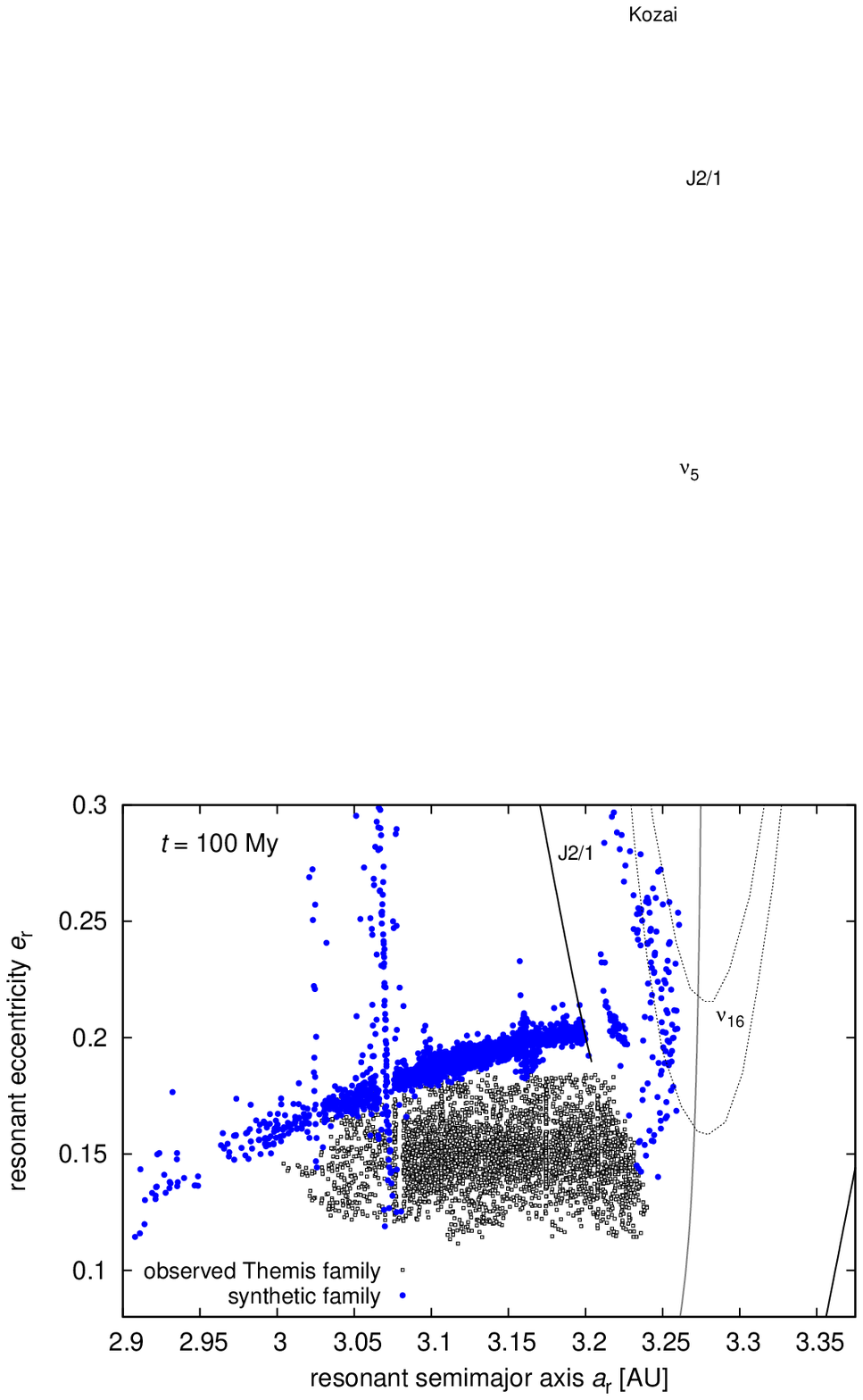}
	\caption{A simulation of a synthetic Themis-like family formation event.
	The figure represents the state $100\,\mathrm{Myr}$ after the breakup.
	The orbital distribution of observed Themis family is displayed
	in the proper elements (black squares), while the orbital
	distribution of the synthetic family is displayed in
	the resonant elements (blue circles). This causes the mutual
	shift in eccentricity.
	Note the differences in extent of both families and the 
	eccentricity dispersion beyond the 11:5 mean-motion
	resonance with Jupiter.
        }
        \label{fig:themis}
\end{figure}

(ii)~At the same time, it is required that the breakup takes place
when the true anomaly $f_{\rm imp} \simeq 0^\circ$
for this ejection scenario to work; otherwise,
the number of objects landing in the islands decreases as well.
But this particular impact
geometry does {\em not\/} seem to be compatible with the observed
shape of Themis family, namely with a large eccentricity dispersion
of the family below the J11/5 resonance, at $a_{\rm p} = 3.03\,{\rm AU}$
(see Fig.~\ref{fig:themis}).

(iii)~Starting with a more reasonable size-dependent velocity field
-- as used in \cite{Broz_Morbidelli_2013Icar..223..844B} for Eos family,
which has a similar parent body size -- makes the contribution
to the long-lived population as low as $N(D > 5\,{\rm km}) = 2$.

(iv)~There is no way to explain the existence of asteroids
in island A by the ejection, both the eccentricities and
inclinations of the observed A-island objects are too large.
Therefore, the ejection from Themis is not a viable hypothesis
in case of island A.

\bsp

\label{lastpage}

\end{document}